# Lattice-guided growth of dense arrays of aligned transition metal dichalcogenide nanoribbons with high catalytic reactivity

Z. Ma,[1] P. Solís-Fernández,[2*] K. Hirata,[3] Y.-C. Lin,[4,5] K. Shinokita,[6] M. Maruyama,[7] K. Honda,[3] T. Kato,[8] A. Uchida,[2] H. Ogura,[8] T. Otsuka,[8,9,10,11,12] M. Hara,[13,14] K. Matsuda,[6] K. Suenaga,[5] S. Okada,[7] T. Kato,[8,10] Y. Takahashi,[3,15] & H. Ago[1,2,16*]

[1] Interdisciplinary Graduate School of Engineering Sciences, Kyushu University, Japan.

[2] Faculty of Engineering Sciences, Kyushu University, Fukuoka 816-8580, Japan

[3] Department of Electronics, Graduate School of Engineering, Nagoya University, Japan.

[4] Nanomaterials Research Institute, National Institute of Advanced Industrial Science and Technology (AIST), Tsukuba 305-8565, Japan.

[5] The Institute of Scientific and Industrial Research, Osaka University, Japan.

[6] Institute of Advanced Energy, Kyoto University, Kyoto, 611-0011, Japan.

[7] Department of Physics, Graduate School of Pure and Applied Sciences, University of Tsukuba, Tsukuba 305-8571, Japan.

[8] Graduate School of Engineering, Tohoku University, Sendai 980-8579, Japan.

[9] Research Institute of Electrical Communication, Tohoku University, Sendai 980-8577, Japan.

[10] Advanced Institute for Materials Research (AIMR), Tohoku University, Japan.

[11] Center for Science and Innovation in Spintronics, Tohoku University, Japan

[12] Center for Emergent Matter Science, RIKEN, Saitama 351-0198, Japan

[13] Faculty of Advanced Science and Technology, Kumamoto University, Japan.

[14] Institute of Industrial Nanomaterials, Kumamoto University, Japan.

[15] WPI Nano Life Science Institute (WPI-NanoLSI), Kanazawa University, Japan.

[16] Center for Semiconductor and Device Ecosystem (CSeDE), Kyushu University, Japan.

*Corresponding authors: Pablo Solís-Fernández (solisfernandez.pablo.kyudai@gmail.com), Hiroki Ago (ago.hiroki.974@m.kyushu-u.ac.jp)

Transition metal dichalcogenides (TMDs) exhibitbelow 10 nm and longitudinal axis parallel to the unique properties and potential applications whenzigzag direction, being also extensible to the growth reduced to one-dimensional (1D) nanoribbons (NRs),of $WS_2$ NRs and $MoS_2$-$WS_2$ hetero-nanoribbons. owing to quantum confinement and high edgeGrowth is influenced by both substrate and CVD densities. However, effective growth methods fortemperature, indicating the role of anisotropic self-aligned TMD NRs are still lacking. Weprecursor diffusion and substrate interaction. The 1D demonstrate a versatile approach for lattice-guidednature of the NRs was asserted by the observation of growth of dense, aligned $MoS_2$ NR arrays via chemicalCoulomb blockade at low temperature. Pronounced vapor deposition (CVD) on anisotropic sapphirecatalytic activity was observed at the edges of the substrates, without tailored surface steps. ThisNRs, indicating their promise for efficient catalysis. method enables the synthesis of NRs with widths

## Introduction

Transition metal dichalcogenides (TMDs), a family of two-dimensional (2D) semiconducting materials, have been gaining interest for their potential in post-Si electronic devices to extend Moore's law[1,2]. In particular, TMDs are expected to play an important role in future logic circuits, such as gate-all around (GAA) and nanosheet transistors, offering ultrathin 2D channels with high carrier mobility, in contrast to conventional Si electronics[3]. Such applications require the use of narrow TMD channels with atom-level thickness, with nanoribbons (NRs), one-dimensional (1D) forms of 2D materials, fulfilling this requirement. Moreover, reducing the dimensionality from 2D to 1D substantially alters the electronic structure of the materials, leading to unique physical properties, such as edge- and width-dependent band structure[4], metallic edge states[5], magnetic ordering[5,6], piezoelectricity[7], and robust carrier mobility[8]. Furthermore, the high edge-to-area ratio of NRs can enhance their catalytic activity[9,10].

TMD NRs are obtained via two main approaches, namely, top-down and bottom-up methods. Top-down approaches typically rely on lithographic techniques[11,12], allowing position-selective patterning but suffering from low throughputs and difficulties in controlling the edge structures, and with ribbon widths limited by the resolution of lithography and etching techniques. Some studies have used these methods to create TMD nanoribbon-based field-effect transistors (FET), providing initial insights into their electronic properties[13,14] and proving their 1D nature[15]. Chemistry-based top-down methods offer higher throughput but introduce a large amount of defects[16], while mechanical exfoliation from bulk materials can produce aligned NR arrays but lacking control over width and thickness[17]. Chemical vapor deposition (CVD) is a promising bottom-up approach for large-scale synthesis of TMD NRs with controlled width and orientations. However, CVD typically yields triangular 2D grains with zigzag (ZZ) edges [4,18], with only limited studies on the synthesis of TMD NRs by CVD[19–23]. Vapor-solid-liquid CVD growth of $MoS_2$ NRs has been demonstrated using metal alloy and Ni nanoparticles[19,20]. Additionally, $MoS_2$ NRs have been produced by bottom-up synthesis using step-terrace structures specifically formed on surfaces like Au[23,24] and β-$Ga_2O_3$[22] that serve as nucleation points and assist the growth to form ribbon-like structures. However, this method relies on very specific substrates with special miscut angles to obtain the tailored steps. Furthermore, precise control over the structure of the NR edges is still challenging in this step-templated growth mode, often leading to saw-tooth NR edges[22]. Hence, there is a pressing need for a more versatile approach to produce narrow, high-density aligned TMD NRs.

Here, we demonstrate a self-aligned growth mode of TMD NRs by CVD, leveraging lattice guidance on low-symmetry sapphire ($\alpha$-$Al_2O_3$) substrates. Such kind of substrates have been previously used to grow aligned carbon nanotubes[25,26] and graphene NRs[27], and enabled growing aligned rectangular grains of $MoS_2$ and $WS_2$[28]. The anisotropic surface atomic arrangements of a- and r-plane sapphire substrates control the growth directions of $MoS_2$, yielding narrow and highly aligned NR arrays. This alignment is particularly interesting as it paves the way for potential industrial scaling of the growth process, a crucial step towards practical applications of these NRs. Temperature and a proper exposure to the precursors, in addition to the substrate, were found to be key parameters for 1D growth, with density functional theory (DFT) calculations hinting to the role of anisotropic precursor diffusion. The NRs display remarkable carrier mobility, even at narrower widths, underscoring their exceptional quality[8], with the observation of Coulomb blockade at low temperatures confirming the 1D nature of the NRs[15]. The method can be also developed to the synthesis of $MoS_2$-$WS_2$ hetero-NRs. Furthermore, we revealed that edges of $MoS_2$ NRs act as highly active catalyst sites for hydrogen evolution reaction (HER). Our findings offer an original pathway to fabricate large-scale, aligned TMD NR arrays for catalysis and electronics.

## Results

## Self-aligned growth of TMD NRs

The symmetry of crystals can be used to grow low dimensional materials with controlled shapes and alignments[27,29–31]. Here we introduce a substrate lattice-guided method to produce aligned TMD NRs on a-plane sapphire $(11\bar{2}0)$, which has a two-fold symmetry. The growth is tuned by the temperature and controlled exposure to precursors (sulfur, and $MoO_3$ or $WO_3$, to grow $MoS_2$ or $WS_2$, respectively) during the CVD (Fig. 1A). Further experimental details are described in the "Materials and Methods" section. Fig. 1B and Fig. S1 show scanning electron microscope (SEM) images of regions with dense arrays of aligned $MoS_2$ NRs, grown at 1100 °C on a-plane sapphire. The selection of the substrate is critical to produce NRs, with the same growth conditions producing triangular flakes on c-plane sapphire (Fig. S2). As will be shown later, the NRs are oriented along the $[1\bar{1}00]$ direction of the sapphire substrate. Here, the choice of substrate, growth temperature and amount of precursor are all indispensable for the successful growth of NRs. These ribbons can have lengths up to a few tens of μm and widths that are below 10 nm. The distribution of the width and length of ~150 NRs is shown in Fig. 1C. This data, collected from the SEM images in Fig. S3, evidences a positive correlation between the length ($L$) and width ($W$) of the NRs, with aspect ratios ($L/W$) of up to ~80. This choice of SEM imaging here represents a balance between the greater accuracy of other microscopic techniques and the high throughput of SEM, although it prevents the inclusion of both the narrowest and longest NRs. To delve deeper into the characteristics of the narrowest NRs, atomic force microscope (AFM) images of the NRs were taken, as displayed in Fig. 1, D and E. The width of these NRs and the AFM image of the whole area are shown in Fig. S4. The NRs exhibited apparent widths down to ~24 nm by AFM. But as will be discussed afterwards, scanning transmission electron microscopy (STEM) images revealed the existence of NRs narrower than 10 nm. This suggests that the widths measured by AFM are likely overestimated due to the effect of tip convolution.

The single-layer nature of the NRs was confirmed by AFM (Fig. S4, B and C), with a measured average thickness of ~4.5 Å, and further supported by optical measurements. The observation of the Raman shift of the characteristic $A_{1g}$ and $E_{2g}$ peaks of $MoS_2$, along with their relative distance (~20 $cm^{-1}$), indicates that the NRs are single-layered (Fig. S5)[32]. Photoluminescence (PL) mappings were collected from a representative area populated with NRs (Fig. 1, F-I), as well as from a distinct area that contained both triangular $MoS_2$ flakes and NRs (Fig. S6). The mappings reveal a lower PL intensity for the NRs compared with the triangular flakes. This weaker PL of the NR can be attributed to the larger laser spot size (~1 μm) relative to the NR width, rather than the NR being multilayer. This is also supported by the observation that narrower NRs tend to exhibit lower PL intensities (Fig. 1, G and I). The PL energy is also shifted to higher energies for the NRs, which can be attributed to a higher strain from the substrate coupled to the quantum confinement effect arising from the reduced dimensionality of the nanoribbons [33]. The narrower PL of the NRs (Fig. S6D) suggests higher crystalline quality compared to the triangular flakes[34].

The orientation of the NRs was determined using polarization-resolved second-harmonic generation (SHG), a technique that can probe the structural symmetry of TMDs[17,35]. The parallel polarization component of the SHG signal reaches a maximum intensity when the polarization of the incident laser aligns with the armchair direction of the $MoS_2$[35]. Therefore, the SHG signal of the NRs, depicted in Fig. 1J, indicates that the longitudinal axis of the NRs is parallel to the $MoS_2$ ZZ direction. Interestingly, instead of the 6-fold symmetric signal expected from $MoS_2$ flakes, we observed a decrease in the intensity of the lobes in the direction perpendicular to the longitudinal axis of the NRs. This anomaly indicates a deviation from the usual single $\cos^2$ component used to fit the SHG signal of $MoS_2$ flakes[35]. A more accurate fit for the NRs is achieved with a product of two $\cos^2$ components. One of these components accounting for the 6-fold symmetry of the $MoS_2$ lattice, and another for the 2-fold symmetry (Fig. S7). Small asymmetries of the lobes in exfoliated TMD NRs have recently been attributed to strain along the perpendicular direction of the NR axis[17]. Additionally, these asymmetries can also be ascribed to the 1D nature of the NRs[36].

## Analysis of edges and orientation of the NRs

The atomic structure of the NR and their edges, as well as their orientation with respect to the sapphire substrate, were characterized by STEM. Fig. 2A displays a top-view STEM image of a ~5.7 nm-wide NR supported on single-layer graphene (SLG), with the inset showing the fast Fourier transform (FFT) of the image. The STEM analysis confirms that the NR consists of a single layer with a 1H structure and indicates that the longitudinal axis of the NRs is aligned parallel to the ZZ direction of the $MoS_2$ lattice[4,23]. These observations are consistent with the AFM and SHG results discussed previously. A high-resolution image of the center of the NR is presented in Fig. 2B, in which the absence of defects reflects the high crystalline quality of the NRs.

An examination of the edges of the NRs reveals differences between the two sides of the ribbon. As seen in the model overlapped in Fig. 2B, the right and left edges of the ribbon are expected to be terminated by Mo (ZZ-Mo) and $S_2$ (ZZ-$S_2$) atoms, respectively. Fig. 2C shows that the ZZ-Mo edge is relatively straight, and a closer inspection shows that the Mo atoms at the edge are terminated by single S atoms (ZZ-Mo(S)) (see Fig. S8, A-C). The S termination of the ZZ-Mo edge has been previously reported for the edges of $MoS_2$ nanoclusters[37], and suggests a sulfur-rich environment during the growth of the NRs[38]. In contrast, analysis of the edge on the ZZ-$S_2$ side of the NR reveals a reconstruction of the edge structure (Fig. S8D). The ZZ-$S_2$ edge maintains a global direction that is parallel to the ZZ direction of $MoS_2$, although it is locally terminated with a mixture of both ZZ-$S_2$ and ZZ-Mo(S) sections forming angles of 60° between them. This is in accordance with calculations showing that ZZ-Mo(S) edges are expected to be more stable than ZZ-$S_2$ edges in S-rich growth environments[39,40]. A similar behavior has been reported for the case of $MoSe_2$ NRs, for which the Mo edge is smoother than the Se edge[41]. This reconstruction can be attributed to the higher stability of the ZZ-Mo(S) sections compared to the ZZ-$S_2$ under the actual growth conditions in a sulfur-rich environment[38].

The alignment direction of the NRs with respect to the sapphire substrate was determined from cross-sectional high-resolution images of a 26 nm wide NR sectioned perpendicularly to its longitudinal axis. A wide transmission electron microscopy (TEM) view of the NR can be seen in Fig. 2D, with a detailed STEM image of the ZZ-Mo(S) edge and the substrate shown in Fig. 2E. The elemental composition of the NR measured by electron energy loss spectroscopy (EELS) mappings indicates the presence of $MoS_2$ on the top of the sapphire surface (Fig. 2F). Additionally, EELS spectra of the $MoS_2$ (Fig. S9) show the characteristic S and Mo edges. The sapphire crystal face shown in the cross-section images is representative of the m-plane[18], proving that the NRs grow parallel to the $[1\bar{1}00]$ direction. This observation confirms the lattice-oriented growth of the NRs, which can be attributed to the linear atomic arrangement of the O-atoms in the surface of the a-plane for each of the two O layers (Fig. S10). The factors enabling the linear growth of the NRs is discussed in the next section.

## Role of growth temperature and substrate

We found that the growth temperature is a critical factor in producing NRs, observing a transition from the growth of randomly oriented $MoS_2$ flakes to aligned NRs as the temperature increased. As depicted in Fig. 3A, sub-micron sized, randomly oriented $MoS_2$ flakes are obtained at lower temperatures (750 °C). As will be shown below, the small size of the flakes is related with a controlled exposure to the precursors, which plays an essential role in the growth of NRs. When the temperature is raised to 950 °C, the flakes begin to show an alignment along the $[1\bar{1}00]$ direction of sapphire, and their shape transitions from the triangles to irregular grains (Fig. 3B). At 1100 °C, sharp and narrow NRs are produced (Fig. 3C). This temperature dependence highlights the importance of the anisotropic diffusion of intermediate species on the a-plane surface. At high temperatures, the surface diffusion is enhanced, resulting in the formation of $MoS_2$ NRs. However, further increasing the temperature results in an increased amount of precursors on the sapphire surface. Hence, increasing the temperature up to 1200 °C did not produce notably narrower NRs (Fig. 3D). Instead, it increased the amount of $MoO_x$ species on the sapphire substrate, producing thicker NRs, a higher density of NRs, and the formation of thick Mo-based byproducts in several areas of the substrate (Fig. S11). At these high temperatures it was also evident the growth of some NRs along directions different to $[1\bar{1}00]$.

Previous research highlighted the role of temperature in transitioning from the growth of triangular to rectangular-shaped $MoS_2$ flakes on a-plane sapphire by face-down exposure to the precursors[28]. However, the high temperatures (~1100 °C) necessary to grow NRs led to enhanced sublimation of $MoO_3$, resulting in an uncontrolled growth and the formation of thick Mo-based materials (Fig. S12A). To decrease the width of the $MoS_2$ flakes and successfully produce NRs, it was essential to increase the CVD temperature while limiting the amount of $MoO_x$ species on the sapphire substrate. This was achieved by positioning the sapphire downstream and far from the precursors (Fig. S12B).

Despite the presence of steps on the a-plane sapphire surface, attributed to minor miscut angles, their influence on the growth and alignment of the $MoS_2$ NRs was found to be negligible. This is evident from the AFM images, which show a misalignment of approximately 13° between the NRs and the steps (Fig. 3E). This observation confirms that the 1D growth is governed by the lattice structure of the sapphire, rather than the surface steps. To confirm the role of the low symmetry of the substrate, the growth of $MoS_2$ NRs was also performed on r-plane sapphire $(1\bar{1}02)$ [25,27]. The $MoS_2$ NRs grew along the preferential direction $[1\bar{1}0\bar{1}]$ of the sapphire, rather than following the direction of the sapphire steps (Fig. 3F and Fig. S13). In contrast, the growth of TMD NRs on other substrates, such as faceted Au, is reported to be initiated and guided by steps on the surface[23]. The lattice-guided growth ensures a more uniform and predictable alignment of the NRs, which is critical for the reproducibility and scalability of the synthesis. Additionally, lattice-guided growth is less dependent on the surface morphology of the substrate, potentially allowing for a wider range of substrate choices and simplifying the preparation process.

DFT calculations were performed to understand the relationship between the aligned 1D growth and the crystal lattice of the substrate. A small, NR-like $MoS_2$ flake with a trapezoidal shape and ZZ edges was positioned atop an O-terminated a-plane sapphire lattice. A simplified model of the structure containing only the top layer of O atoms of the sapphire is shown in Fig. 3G, with examples of the whole models shown in Fig. S14. The total energy was computed across a range of relative positions of the flake over the sapphire lattice (Fig. 3H and Supplementary Movie S1). The calculations revealed that the relative position between the Mo atoms in the $MoS_2$ flake and the surface atoms of the sapphire is important for the stability of the system. Minimum energies were achieved when the Mo atoms were in close proximity to the O atoms of the sapphire, in which a substantial hybridization between Mo 4d and O 2p orbitals occurred. This strong interaction between the Mo atoms and the O atoms stabilizes the system and lowers the total energy.

Furthermore, the calculations indicated that the barrier energies for the diffusion of the flakes are smaller in the $[1\bar{1}00]$ direction of the sapphire. On the other hand, the barrier in the $[0001]$ direction (perpendicular to $[1\bar{1}00]$) is more than five times larger (see Fig. 3, I and J). These directions correspond to those in which the number of Mo-O pairs remain almost constant ($[1\bar{1}00]$) and to that where there is a substantial modulation in the atomic arrangement between the two lattices ( $[0001]$ ). A preferential diffusion of $S_2$ monomers along the $[1\bar{1}00]$ direction of the a-plane has also been recently reported[31], leading to the preferential growth of $WS_2$ trapezoidal flakes along such direction, while similar anisotropic diffusion barriers have been found for the aligned growth of $WS_2$ on m-plane quartz[42]. This suggests that precursor diffusion is more favorable in the $[1\bar{1}00]$ direction, resulting in preferred growth along this axis.

## Growth of hetero-nanoribbons

The versatility of our method allows for the growth of other kinds of NRs, such as $WS_2$ NRs, as shown in Fig. S15, through a simple change of the precursor. By adjusting the composition of the precursor during the growth or at a subsequent growth step, the lattice-guided CVD allows to grow hetero-nanoribbons (hNRs) composed of lateral heterostructures. These hNRs are of considerable interest due to their potential in advanced electronic and optoelectronic devices[43]. The seamless stitching within a single atomic layer in these lateral heterostructures offers opportunities for engineering of the band structure and their properties, making them promising candidates for applications in nanoelectronics and optoelectronic applications[2,44,45]. The ability to grow such structures in an aligned manner further enhances their potential for scalable device integration.

Figs. 4A and 4A depict SEM and AFM images of isolated in-plane $MoS_2$-$WS_2$ hNRs, respectively, which were obtained by switching the precursor from $MoO_3$ to $WO_3$. In both images, two distinct regions can be discerned: one at the center of the hNR and the other surrounding it. The photoluminescence (PL) mapping collected in one of such hNRs clearly shows that the center is composed of $MoS_2$, with the surrounding material being $WS_2$ that laterally extended from the original $MoS_2$ seed during the second stage (Fig. 4, C and D). The presence of these distinct peaks, without any intermediate peaks characteristic of $Mo_xW_{(1-x)}S_2$ alloys, proves the formation of a lateral heterojunction similar to those observed for large flakes[44–46]. Individual PL spectra collected at the distinct regions of the hNR ($MoS_2$, $WS_2$ and the interface) are shown in Fig. 4E, while the evolution of the spectra along two different directions is shown in Fig. S16. Given that the components of the hNR are smaller than the size of the PL laser spot (~1 μm), our setup does not allow for a strict differentiation of pure $MoS_2$ and $WS_2$ areas within a single spectrum. Only pure $WS_2$ can be discerned at the ends of the ribbon for hNRs longer than a few microns (top and bottom areas of the hNR in Fig. S16D). The Raman mapping of a broader hNR, shown in Fig. S17, allows us to observe a sharp transition between the regions with $WS_2$ peaks and with $MoS_2$ peaks. This, along with the distinct PL emissions observed in Fig. S16, proves the formation of well-defined $MoS_2$ and $WS_2$ domains within the hNRs, rather than the formation of alloyed regions[46]. To further characterize the $MoS_2$-$WS_2$ interface at the atomic scale, we performed high-resolution STEM imaging (Fig. 4F and Fig. S18). The results revealed a narrow diffusion region of approximately 6 nm at the interface, in which primarily W atoms (brighter in the STEM images) diffused into the $MoS_2$ lattice, forming substitutional defects. This limited interdiffusion is consistent with our growth process, in which $MoS_2$ is grown first. Interestingly, the diffusion region tends to be narrower at the smoother ZZ-Mo(S) edge of the $MoS_2$ NRs, compared to the ZZ-$S_2$ edge (Fig. S18A-B). The edge-dependent behavior suggests that refining growth conditions for both the $MoS_2$ and $WS_2$ components could potentially minimize interdiffusion, leading to even sharper interfaces. For instance, this could be achieved by producing straighter edges of the initial $MoS_2$ NR or by decreasing the processing temperature for the subsequent growth of $WS_2$. Despite this localized mixing at the atomic scale, the overall structure maintains distinct $MoS_2$ and $WS_2$ domains, as evidenced by PL and Raman, and by EELS (Fig. S18C) of the different regions. Importantly, the STEM analysis also revealed that the $WS_2$ domain is single-crystalline and maintains the same crystallographic orientation as the $MoS_2$ domain, indicating a seamless stitching between the two materials.

SEM imaging and PL mappings of wide-areas (Fig. 4, G-H and Fig. S19) demonstrate that the growth of the $MoS_2$-$WS_2$ hNRs is uniform across the sapphire substrate, with all the original $MoS_2$ NRs acting as seeds to form the in-plane hNRs. The proposed method can be employed to produce A-B-A (with A, B: $MoS_2$, $WS_2$, $MoSe_2$ …) heterojunctions along the longitudinal axis of the hNR, and with nanometer-sized widths determined by the initial width of the NR acting as seed[45,47].

In summary, the growth of hNRs proves that our method allows for precise control over both dimensionality and composition, resulting in truly one-dimensional TMD heterostructures. This opens up the possibility to design complex nanostructures with tailored properties.

## Edge-dominated HER catalytic activity

TMDs are known as promising catalysts for HER, with reactivity being higher at the edges compared to the basal planes[48]. Due to their high edge-to-area ratio, NRs are expected to be ideal candidates for high HER activity compared to 2D flakes. To investigate the local HER activity of $MoS_2$ NRs, we employed high-resolution scanning electrochemical cell microscopy (SECCM)[48]. This technique enables the spatial investigation of the electrochemical activity, distinguishing the activity between the edges and central regions of the NRs and offering insights into the spatial distribution of catalytic performance.

To conduct HER measurements, the $MoS_2$ NRs were transferred onto highly oriented pyrolytic graphite (HOPG) substrate, which serves as a conductive but catalytically inactive substrate. Fig. 5A shows a SEM image of two wide $MoS_2$ NRs on HOPG, with widths of ~290 nm and ~250 nm, and of a narrow NR with a width of ~55 nm. Fig. 5B displays the corresponding HER current mapping image for the same area, measured at -0.95 V vs. RHE (reversible hydrogen electrode). The HER current mapping of the wider NRs distinctly reveals the current from each edge. Remarkably, the current from the edges is much higher than that from center of the NRs. As shown in the current profile in Fig. 5C, the current from edges is approximately two orders of magnitude higher than that at the center indicating a superior catalytical activity of the edges.

Fig. 5, D and E also show that the edges of the NRs have a lower overpotential and smaller Tafel slope compared to the center of the NR and to the HOPG. The Tafel slope provides insight into the rate-determining step of the HER, with smaller slopes indicating better catalytic performance. Our estimated Tafel slope of 139 mV/dec for the $MoS_2$ NR edges suggest that the Volmer reaction is likely the rate-determining step in our system, indicating that hydrogen adsorption is the most favorable reaction for our $MoS_2$ NRs. To contextualize our results within the broader field of HER catalysts, we compared our findings with those of other materials (see Table S1). Interestingly, the observed wide NRs evidence differences in the current contribution from each edge (see, for example, the rightmost NR in Fig. 5, B and C), suggesting that one edge generates more current than the other. This observation implies inherent disparities in the edge morphology and/or composition between the two edges of a same NR (as discussed in Fig. S8), and their impact on HER activity. Current studies are underway to elucidate the specific effects of edge morphology and composition on the HER performance.

Although the SECCM used in this study is capable to resolve features down to several tens of nanometer, it presents a challenge distinguishing between the edges of the narrow NR (~ 55 nm). This is because the width of the NR is comparable to the diameter of the nanopipette used in the SECCM measurements (~ 40 nm). Consequently, the HER current response measured on the narrow NR is roughly equal to twice that on each edge of the wide NR because, as it encompasses the HER current responses from both the left and right edges of the narrow NR. This highlights the efficiency of having dense arrays of narrow NRs, with their high edge-to-area ratio, for increasing the current density in HER applications compared to large 2D flakes.

## Transport properties and performance of MoS2 NR field-effect transistors

The transport properties of the $MoS_2$ NRs were evaluated using back-gated FETs, as shown in the SEM image of Fig. 6A. The transfer characteristics of a FET with a 110 nm-wide NR channel is shown in Fig. 6B, displaying a carrier mobility of 44 $cm^2V^{-1}s^{-1}$. The device sustains a current density of approximately 22.7 $\mu A\mu m^{-1}$ (3.25 $MAcm^{-2}$) at a bias voltage of 1 V, which is comparable to the values obtained for high-quality exfoliated $MoS_2$ flakes[49]. This high current density, along with the remarkable mobility, underscores the high quality of the $MoS_2$ NRs and their potential for applications in high-performance electronic devices[8]. The linearity of the output characteristics of the device, shown in Fig. S20, evidences a good ohmic contact between the NR and the electrodes at room temperature. Fig. 6C presents the carrier mobility versus the on/off ratio for devices with different channel widths, represented by the color of the markers. Although there is a positive correlation between the mobility and the on/off ratio, the mobility appears to be independent of the NR width, with an average value of 25.4 $cm^2 V^{-1} s^{-1}$ at room temperature.

In order to verify the emergence of one-dimensional characteristics in the $MoS_2$ NRs, detailed electrical conductivity measurements were conducted at low temperature (4.2 K) for a NR with width and length of 30 nm and 300 nm, respectively. This device is small enough to behave as a quasi-zero-dimensional quantum dot (QD). If defect-free $MoS_2$ NRs are synthesized, we can expect to observe Coulomb oscillation with fine oscillation periods, where the entire $MoS_2$ NR becomes a single QD. On the other hand, in the case of $MoS_2$ NRs with many defects, multiple QDs will be dominant, and it can be expected that only non-periodic Coulomb oscillation will be observed. In other words, the quality of the $MoS_2$ NRs can be inferred from the shape of the observed quantum features. Since the threshold of the n-type current at this low temperature exists around $V_G$ from 55 to 60 V (Fig. S21A), the $V_D$-$V_G$ dependence of the differential conductance ($dI_D/dV_D$) was measured in this region. As a result, large periodic Coulomb oscillations with peak intervals $\Delta V_G$ ranging from 1.23 to 1.4 V (#L1, #L2) (Fig. 6D) and Coulomb diamonds (Fig. S21B) were observed. Closer examination of a single Coulomb oscillation, specifically within the $V_G$ range of 56-57 V, uncovered the presence of finer periodic oscillation components (Fig. S21C). Detailed measurements in this expanded region revealed the existence of eight highly periodic Coulomb diamonds (#S1 to #S8) as shown in Fig. 6E.

The observation of Coulomb diamonds with both long (#L1, #L2) and short periods (#S1 to #S8) indicates the presence of QD-like structures and QDs of different sizes within the synthesized NR. Although it is difficult to completely identify the QD sizes due to the presence of multiple structures, order estimation of QD size can be possible from these periodic peaks to infer the origins of the large (#L1, #L2) and small (#S1 to #S8) periodic components. The QD size was calculated based on the general planar model[50]. Since the width of the NR (30 nm) is sufficiently small compared to the $SiO_2$ oxide film (300 nm), the electric field concentration effect was also taken into account in the calculations[51]. The results showed that the areas for #L1 and #L2 are 100-116 nm², while those for #S1 to #S8 are approximately 1470-3390 nm² (Fig. 6F). If we consider the entire nanoribbon (with dimensions 30 nm × 300 nm) acting as a single QD, the QD area would be approximately 9000 nm². The QD area obtained for #S1 to #S8 corresponds to about 16.3-37.6% of the entire $MoS_2$ NR, suggesting the formation of relatively large QDs across the entire $MoS_2$ NR. Considering edge roughness and the Schottky barrier with the electrodes, it is expected that the effective QD area is smaller than the geometrically calculated area (9000 nm²). Therefore, this explanation of the entire $MoS_2$ NR structure acting as a single QD may be reasonable by considering demonstrated the high-quality of the $MoS_2$ NRs (Fig. 2, A-C). On the other hand, the long-period components (#L1, #L2) correspond to QD-like structures with a diameter of about 10 nm, and are likely due to minor local defects or edge roughness present in the $MoS_2$ NR.

## Conclusion

In this work, we demonstrate the lattice-guided mode for the self-aligned growth of TMD NRs on low-symmetry substrates, such as a-plane sapphire. DFT calculations revealed that this method is enabled by the anisotropy in the precursor diffusion on the a-plane sapphire that results in the alignment of the NRs along the $[1\bar{1}00]$ direction of sapphire. By increasing the growth temperature up to 1100 ºC, the shape of the $MoS_2$ evolved from the normal triangular flake shape to NRs with widths below 10 nm. SHG measurements and optical and STEM images revealed that the longitudinal axis of the NRs is parallel to the $MoS_2$ zigzag direction. We showed that our method can be extended to the growth of lateral NR heterostructures by varying the precursors during the growth process. Furthermore, the edges of the $MoS_2$ NRs showed a high HER current response, which, besides to their high edge-to-area ratio, makes them promising candidates for efficient catalysis applications. Both STEM and transport property analysis evidenced the high quality of the NRs, highlighting their potential for future electronic and optoelectronic devices.

## Materials and Methods

### Synthesis of the NRs

The growth of the TMD NRs was conducted in a 4-zone furnace with a 26 mm inner diameter quartz tube at ambient pressure with a flow of 300 SCCM of Ar as the carrying gas (Fig. S12). Sulfur (Mitsuwa Chemical Co., 99.999%, 120 mg) and $MoO_3$ (Wako Pure Chemical Industries, 99.9%, 15 mg) powder precursors were located upstream and heated at 200 °C and 600 °C, respectively. For the growth of $WS_2$, the $WO_3$ precursor was heated at 1060 °C. The a-plane sapphire substrate (Kyocera Co.) used had a small miscut angle (<0.3°) off the ideal ($11\bar{2}0$) plane, resulting in the formation of step edges that are determined by the magnitude and direction of the miscut angle, rather than following a specific crystallographic direction of the sapphire. The sapphire was cleaned by sonication in acetone and isopropanol prior to the CVD growth. Additional experiments were also performed using r-plane sapphire (Kyocera Co.). For the effective growth of NRs, we found that limiting the amount of precursors is a critical factor. For this, the sapphire substrate was located downstream, rather than face-down above the $MoO_3$ precursor (see Fig. S12). The growth was performed by heating the sapphire at temperatures ranging from 750 °C to 1200 °C, with 1100 °C being the optimal temperature for obtaining narrow NRs. The growth duration at the target temperature was 10 min, followed by cooling of the furnace.

For the growth of the $MoS_2$-$WS_2$ hNRs, we employed a two-step process, to avoid the mixing of the precursors. $MoS_2$ NRs were first produced by the method explained above ($MoO_3$ and sulfur precursors at 600 °C and 200 °C, respectively, with the substrate at 1100 °C). The as-grown $MoS_2$ NRs were then transferred to a separate, dedicated furnace for the $WS_2$ growth ($WO_3$ and sulfur precursors at 1060 °C and 200 °C, respectively, and the substrate at 1100 °C).

## Transfer of the NRs

The transfer of the NRs to $SiO_2$ and HOPG substrates was performed by a wet-transfer method. First, polymethyl methacrylate (PMMA) was spin-coated on the as-grown NRs. The PMMA with the NRs was released from the sapphire by etching in a 1M KOH aqueous solution for up to 10 minutes. The detached PMMA with the NRs was washed in ultrapure water and deposited in the target substrate ($SiO_2$ or HOPG), which was then heated to 125 °C. Afterwards, the PMMA was removed in hot acetone (60 °C) for 8 hours and dried with a stream of $N_2$.

Graphene-coated Quantifoil TEM grids were used to support the narrower NRs for STEM measurements. SLG was obtained by CVD on Cu thin films sputtered on c-sapphire[52], and transferred to Quantifoil TEM grids without the use of polymers. The SLG/TEM grid was then attached to the as-grown NRs, and immersed in KOH to detach the NRs from the sapphire substrate. After detachment, the NRs/SLG/TEM grid was washed in ultrapure water.

## Characterization

The optical characterization of the samples included confocal Raman and PL spectroscopies, performed at room temperature using a Nanofinder 30 spectrometer (Tokyo Instruments) with a 532 nm laser excitation and a power of 0.1 mW. SHG measurements were conducted using a back-scattering geometry with a pulsed laser (900 nm wavelength, 100 fs pulse width, 80 MHz repetition rate) and a spot size of approximately 1 µm. The reflected SHG signal from the sample was collected by the same objective and passed through a series of optical components to a thermoelectrically cooled high-efficiency CMOS detector. Detection was performed without a monochromator, allowing for the capture of small signals at the expense of wavelength-resolved measurement.

Microscopical inspections included SEM (Hitachi S-4800) and AFM (Bruker Nanoscope V) to characterize the NRs. STEM and EELS analyses were performed at room temperature using a JEOL-tripleC#3 ultrahigh vacuum microscope (JEOL-ARM200F-based) equipped with a JEOL delta corrector and a cold field-emission gun operating at 60 kV. EELS core loss spectra were collected by using Gatan GIF Continuum with Rio CMOS camera optimized for low voltage operation. The STEM simulated images were generated using the software Dr. Probe[53].

## DFT calculations

All calculations were based on DFT[54] as implemented in the Simulation Tool for Atom TEchnology (STATE) program package[55]. We used the vdW-DF2 with the C09 functional to describe the exchange-correlation potential energy among the interacting electrons, which enabled the investigation of the weak dispersive interaction between the $Al_2O_3$ surfaces and $MoS_2$ flakes[56]. Ultrasoft pseudopotentials generated by the Vanderbilt scheme were used to describe the interaction between electrons and nuclei[57]. The valence wave functions and deficit charge density were expanded using plane wave basis sets with cutoff energies of 25 and 225 Ry, respectively. The a-plane $Al_2O_3$ surfaces are simulated by a repeated slab model with 8 O atomic layers with a rectangle of $a$=2.598 nm and $b$=1.647 nm. The $MoS_2$ flake is simulated by the smallest trapezoid comprising 10 Mo and 18 S atoms. The atomic structures were fully optimized until the force acting on each atom was less than $1.33 \times 10^{-3}$ Hartree/Bohr. We used the effective screening medium method to investigate the electronic and electrostatic properties of thin films within the conventional DFT framework[58].

## HER measurement

The details of the SECCM setup have been previously reported[48]. Briefly, a moveable nanopipette (tip diameter 40 nm) containing 0.5 M aqueous $H_2SO_4$ electrolyte and an Ag/AgCl quasi-reference counter electrode (QRCE) were used. The applied potential was -0.95 V vs RHE. Potentials were converted to RHE from Ag/AgCl QRCE. The nanopipette was moved to the sample surface at 10 nm/ms using a z-piezo stage and stopped upon detecting a capacitive current (0.3 pA threshold). The redox current was measured after 2 ms with a measurement time of 50 µs. After each measurement, the nanopipette was retracted by 1.0 µm and the sample was moved to the next point in the xy-plane. By repeating this process, topographic and current images were acquired simultaneously. Local cyclic voltammetry (CV) measurements were performed by selecting specific locations in the SECCM image and at a sweep rate of 5.0 V/s.

## FET fabrication and measurement

To assess the transport properties of the NRs, back-gate $MoS_2$ NRs FETs were fabricated on $SiO_2$ (300 nm)/Si. Electron-beam lithography using a double-layer resist of PMMA 495K and PMMA 950K (both 6% by weight in anisole, spin-coated sequentially at 4000 rpm) was used to define the source and drain electrodes, followed by the deposition of Au/Ni (30 nm/3 nm) electrodes by electron-beam evaporation. Room temperature measurements were carried out in vacuum (~$10^{-4}$ Pa) using a Keysight B1500A semiconductor analyzer. The low temperature (~4.2 K) electrical measurements of the $MoS_2$ NR devices were performed with a cryogenic measurement system. The devices were mounted on a dipstick and cooled down by liquid helium. The currents were measured using a semiconductor test system (HP 4142B modular DC source/monitor). The field-effect mobilities were calculated from the equation $\mu_{FE} = g_m \frac{L}{WC_{ox}V_d}$ [52], in which $L$ and $W$ are the channel length and width, respectively, $C_{ox}$ and is the gate capacitance per unit area, and $V_d$ is the drain voltage. The value of the transconductance, $g_m = \frac{\partial I_d}{\partial V_g}$, was obtained from the linear region of the transfer curves.

## Acknowledgments

The authors acknowledge Prof. Yoshiaki Sugimoto for the assistance with AFM measurements.

## Author contributions

Z.M., P.S.-F. and H.A. conceived the idea and designed the experiments. Z.M. and A.U. performed the synthesis of the NRs. Z.M. and P.S.-F. transferred the as-grown NRs to $SiO_2$ and to SLG/TEM grids. Keisuke S. and K.M. performed the SHG measurements. Z.M. and P.S.-F. performed the Raman, PL, SEM and AFM characterization. Kaito H., Kota H. and Y.T. performed the HER measurements. Y.-C.L. and Kazu S. performed the STEM and EELS measurements. S.O. and M.M. performed the DFT calculations. Z.M. and M.H. fabricated the FET devices. Z.M., P.S.-F., Tatsuki K., H.O., T.O. and Toshiaki K. performed the FET measurements and analyzed the results. P.S.-F. and H.A. supervised this project. P.S.-F., M.Z., Toshiaki K. and H.A. wrote the manuscript, with all authors providing input and comments on the manuscript.

## Funding

This work was supported by:

JSPS Grant-in-Aid for Scientific Research on Innovative Areas "Science of 2.5 Dimensional Materials: Paradigm Shift of Materials Science Toward Future Social Innovation" (KAKENHI grant numbers JP21H05232 (to H.A., S.O., and K.M.), JP21H05233 (to S.O. and H.A.), JP21H05235 (to K.M. and Kazu S.), JP22H05478 (to Y.-C.L.), JP22H05441 (to Toshiaki K.), JP22H05459 (to Y.T.), JP24H01165 (to Toshiaki K.), JP24H01189 (to Y.T.), JP24H00478 (to Y.T.)). JSPS KAKENHI grant numbers JP24H00407 (to H.A.), JP23K17863 (to H.A.), JP21K18878 (to H.A.), JP23H00097 (to. Toshiaki K. and T.O.), JP23K17756 (to Toshiaki K.), JP20H05664 (to K.M. and S.O.), JP23H05469 (to K.M. and S.O.), JP22K18986 (to K.M.), JP24K17708 (to Kaito H.). JST CREST grant numbers JPMJCR18I1 (to H.A.), JPMJCR20B1 (to Kazu S. and H.A.), JPMJCR23A2 (to Toshiaki K.), JPMJCR24A5 (to K.M. and S.O.). JST FOREST Program, grant number JMPJFR203K (to Y.T.). JST ACT-X, grant number JPMJAX23DH (to Kaito H.). Cooperative Research Projects of RIEC, Tohoku University. Z.M. acknowledges the financial support from China Scholarship Council (CSC). The computations in this work were performed using the computer facilities at the Supercomputer Center, the Institute for Solid State Physics, the University of Tokyo. The low temperature device measurements were conducted in the Laboratory for Nanoelectronics and Spintronics at Tohoku University.

**Correspondence and requests for materials** should be addressed to Pablo Solís-Fernández and Hiroki Ago.

## References

1. Akinwande, D. *et al.* Graphene and two-dimensional materials for silicon technology. *Nature* **573**, 507–518 (2019).
2. Liu, Y. *et al.* Promises and prospects of two-dimensional transistors. *Nature* **591**, 43–53 (2021).

3. Wang, S. *et al.* Two-dimensional devices and integration towards the silicon lines. *Nat. Mater.* **21**, 1225–1239 (2022).
4. Fu, W. *et al.* Toward edge engineering of two-dimensional layered transition-metal dichalcogenides by chemical vapor deposition. *ACS Nano* **17**, 16348–16368 (2023).
5. Li, Y., Zhou, Z., Zhang, S. & Chen, Z. $MoS_2$ Nanoribbons: High Stability and Unusual Electronic and Magnetic Properties. *J. Am. Chem. Soc.* **130**, 16739–16744 (2008).
6. Cui, P. *et al.* Contrasting structural reconstructions, electronic properties, and magnetic orderings along different edges of zigzag transition metal dichalcogenide nanoribbons. *Nano Lett.* **17**, 1097–1101 (2017).
7. Maruyama, M., Gao, Y., Yamanaka, A. & Okada, S. Geometric structure and piezoelectric polarization of $MoS_2$ nanoribbons under uniaxial strain. *FlatChem* **29**, 100289 (2021).
8. Cai, Y., Zhang, G. & Zhang, Y.-W. Polarity-reversed robust carrier mobility in monolayer $MoS_2$ nanoribbons. *J. Am. Chem. Soc.* **136**, 6269–6275 (2014).
9. Karunadasa, H. I. *et al.* A molecular $MoS_2$ edge site mimic for catalytic hydrogen generation. *Science* **335**, 698–702 (2012).
10. Li, Z., Huang, M., Li, J. & Zhu, H. Large-scale, controllable synthesis of ultrathin platinum diselenide ribbons for efficient electrocatalytic hydrogen evolution. *Adv. Funct. Mater.* **33**, 2300376 (2023).
11. Chen, S. *et al.* Monolayer $MoS_2$ nanoribbon transistors fabricated by scanning probe lithography. *Nano Lett.* **19**, 2092–2098 (2019).
12. Kotekar-Patil, D., Deng, J., Wong, S. L., Lau, C. S. & Goh, K. E. J. Single layer $MoS_2$ nanoribbon field effect transistor. *Appl. Phys. Lett.* **114**, 013508 (2019).
13. Aslam, M. A. *et al.* Single-crystalline nanoribbon network field effect transistors from arbitrary two-dimensional materials. *npj 2D Mater. Appl.* **6**, 1–9 (2022).
14. Canton-Vitoria, R., Hotta, T., Xue, M., Zhang, S. & Kitaura, R. Synthesis and characterization of transition metal dichalcogenide nanoribbons based on a controllable $O_2$ etching. *JACS Au* **3**, 775–784 (2023).
15. Schock, R. T. K. *et al.* Non-destructive low-temperature contacts to $MoS_2$ nanoribbon and nanotube quantum dots. *Adv. Mater.* **35**, 2209333 (2023).
16. Padmajan Sasikala, S. *et al.* Longitudinal unzipping of 2D transition metal dichalcogenides. *Nat. Commun.* **11**, 5032 (2020).
17. Saunders, A. P. *et al.* Direct exfoliation of nanoribbons from bulk van der Waals crystals. *Small* **n/a**, 2403504.
18. Fu, J.-H. *et al.* Oriented lateral growth of two-dimensional materials on c-plane sapphire. *Nat. Nanotechnol.* **18**, 1289–1294 (2023).
19. Li, S. *et al.* Vapour–liquid–solid growth of monolayer $MoS_2$ nanoribbons. *Nat. Mater.* **17**, 535 (2018).
20. Li, X. *et al.* Nickel particle–enabled width-controlled growth of bilayer molybdenum disulfide nanoribbons. *Sci. Adv.* **7**, eabk1892 (2021).
21. Chowdhury, T. *et al.* Substrate-directed synthesis of $MoS_2$ nanocrystals with tunable dimensionality and optical properties. *Nat. Nanotechnol.* **15**, 29–34 (2020).
22. Aljarb, A. *et al.* Ledge-directed epitaxy of continuously self-aligned single-crystalline nanoribbons of transition metal dichalcogenides. *Nat. Mater.* **19**, 1300–1306 (2020).
23. Yang, P. *et al.* Epitaxial growth of inch-scale single-crystal transition metal dichalcogenides through the patching of unidirectionally orientated ribbons. *Nat. Commun.* **13**, 3238 (2022).
24. Xu, H. *et al.* Oscillating edge states in one-dimensional $MoS_2$ nanowires. *Nat. Commun.* **7**, 12904 (2016).
25. Han, S., Liu, X. & Zhou, C. Template-free directional growth of single-walled carbon nanotubes on a- and r-plane sapphire. *J. Am. Chem. Soc.* **127**, 5294–5295 (2005).
26. Ago, H. *et al.* Synthesis of horizontally-aligned single-walled carbon nanotubes with controllable density on sapphire surface and polarized Raman spectroscopy. *Chem. Phys. Lett.* **421**, 399–403 (2006).
27. Solís-Fernández, P., Yoshida, K., Ogawa, Y., Tsuji, M. & Ago, H. Dense arrays of highly aligned graphene nanoribbons produced by substrate-controlled metal-assisted etching of graphene. *Adv. Mater.* **25**, 6562–6568 (2013).
28. Ma, Z. *et al.* Epitaxial growth of rectangle shape $MoS_2$ with highly aligned orientation on twofold symmetry a-plane sapphire. *Small* **16**, 2000596 (2020).
29. Sohn, J. I. *et al.* Direct observation of the structural component of the metal-insulator phase transition and growth habits of epitaxially grown $VO_2$ nanowires. *Nano Lett.* **7**, 1570–1574 (2007).
30. Ago, H., Nakamura, Y., Ogawa, Y. & Tsuji, M. Combinatorial catalyst approach for high-density growth of horizontally aligned single-walled carbon nanotubes on sapphire. *Carbon* **49**, 176–186 (2011).
31. Wang, J. *et al.* Dual-coupling-guided epitaxial growth of wafer-scale single-crystal $WS_2$ monolayer on vicinal a-plane sapphire. *Nat. Nanotechnol.* **17**, 33–38 (2022).

32. Lee, C. *et al.* Anomalous lattice vibrations of single- and few-layer $MoS_2$. *ACS Nano* **4**, 2695–2700 (2010).
33. Liu, Z. *et al.* Strain and structure heterogeneity in $MoS_2$ atomic layers grown by chemical vapour deposition. *Nat. Commun.* **5**, 5246 (2014).
34. Lu, A.-Y. *et al.* Unraveling the correlation between Raman and photoluminescence in monolayer $MoS_2$ through machine learning models. *Adv. Mater.* **34**, 2202911 (2022).
35. Li, Y. *et al.* Probing symmetry properties of few-layer $MoS_2$ and h-BN by optical second-harmonic generation. *Nano Lett.* **13**, 3329–3333 (2013).
36. Ding, Y. *et al.* Transition metal dichalcogenides heterostructures nanoribbons. *ACS Materials Lett.* 1781–1786 (2023) doi:10.1021/acsmaterialslett.3c00314.
37. Lauritsen, J. V. *et al.* Size-dependent structure of $MoS_2$ nanocrystals. *Nat. Nanotechnol.* **2**, 53–58 (2007).
38. Zhao, X. *et al.* Mo-terminated edge reconstructions in nanoporous molybdenum disulfide film. *Nano Lett.* **18**, 482–490 (2018).
39. Zhu, S. & Wang, Q. A simple method for understanding the triangular growth patterns of transition metal dichalcogenide sheets. *AIP Adv.* **5**, 107105 (2015).
40. Cao, D., Shen, T., Liang, P., Chen, X. & Shu, H. Role of chemical potential in flake shape and edge properties of monolayer $MoS_2$. *J. Phys. Chem. C* **119**, 4294–4301 (2015).
41. Chen, Y. *et al.* Fabrication of $MoSe_2$ nanoribbons via an unusual morphological phase transition. *Nat. Commun.* **8**, 15135 (2017).
42. Wang, J. *et al.* Multiple regulation over growth direction, band structure, and dimension of monolayer $WS_2$ by a quartz substrate. *Chem. Mater.* **32**, 2508–2517 (2020).
43. Wang, J., Li, Z., Chen, H., Deng, G. & Niu, X. Recent advances in 2D lateral heterostructures. *Nano-Micro Lett.* **11**, 48 (2019).
44. Gong, Y. *et al.* Vertical and in-plane heterostructures from $WS_2/MoS_2$ monolayers. *Nat. Mater.* **13**, 1135–1142 (2014).
45. Li, M.-Y. *et al.* Epitaxial growth of a monolayer $WSe_2$-$MoS_2$ lateral p-n junction with an atomically sharp interface. *Science* **349**, 524–528 (2015).
46. Huang, Z. *et al.* Broad band modulation of two-dimensional $Mo_{1-x}W_xS_2$ by variational compositions. *J. Phys. D: Appl. Phys.* **57**, 315109 (2024).
47. Zhou, J. *et al.* Morphology engineering in monolayer $MoS_2$-$WS_2$ lateral heterostructures. *Adv. Funct. Mater.* **28**, 1801568 (2018).
48. Takahashi, Y. *et al.* High-resolution electrochemical mapping of the hydrogen evolution reaction on transition-metal dichalcogenide nanosheets. *Angew. Chem. Int. Ed.* **59**, 3601–3608 (2020).
49. Lembke, D. & Kis, A. Breakdown of high-performance monolayer $MoS_2$ transistors. *ACS Nano* **6**, 10070–10075 (2012).
50. Kato, T. *et al.* Scalable fabrication of graphene nanoribbon quantum dot devices with stable orbital-level spacing. *Commun. Mater.* **3**, 1–7 (2022).
51. Lin, Y.-M., Perebeinos, V., Chen, Z. & Avouris, P. Electrical observation of subband formation in graphene nanoribbons. *Phys. Rev. B* **78**, 161409 (2008).
52. Orofeo, C. M. *et al.* Influence of Cu metal on the domain structure and carrier mobility in single-layer graphene. *Carbon* **50**, 2189–2196 (2012).
53. Barthel, J. Dr. Probe: A software for high-resolution STEM image simulation. *Ultramicroscopy* **193**, 1–11 (2018).
54. Kohn, W. & Sham, L. J. Self-consistent equations including exchange and correlation effects. *Phys. Rev.* **140**, A1133–A1138 (1965).
55. Morikawa, Y., Iwata, K. & Terakura, K. Theoretical study of hydrogenation process of formate on clean and Zn deposited Cu(1 1 1) surfaces. *Appl. Surf. Sci.* **169–170**, 11–15 (2001).
56. Cooper, V. R. van der Waals density functional: An appropriate exchange functional. *Phys. Rev. B* **81**, 161104(R) (2010).
57. Vanderbilt, D. Soft self-consistent pseudopotentials in a generalized eigenvalue formalism. *Phys. Rev. B* **41**, 7892–7895 (1990).
58. Otani, M. & Sugino, O. First-principles calculations of charged surfaces and interfaces: A plane-wave nonrepeated slab approach. *Phys. Rev. B* **73**, 115407 (2006).

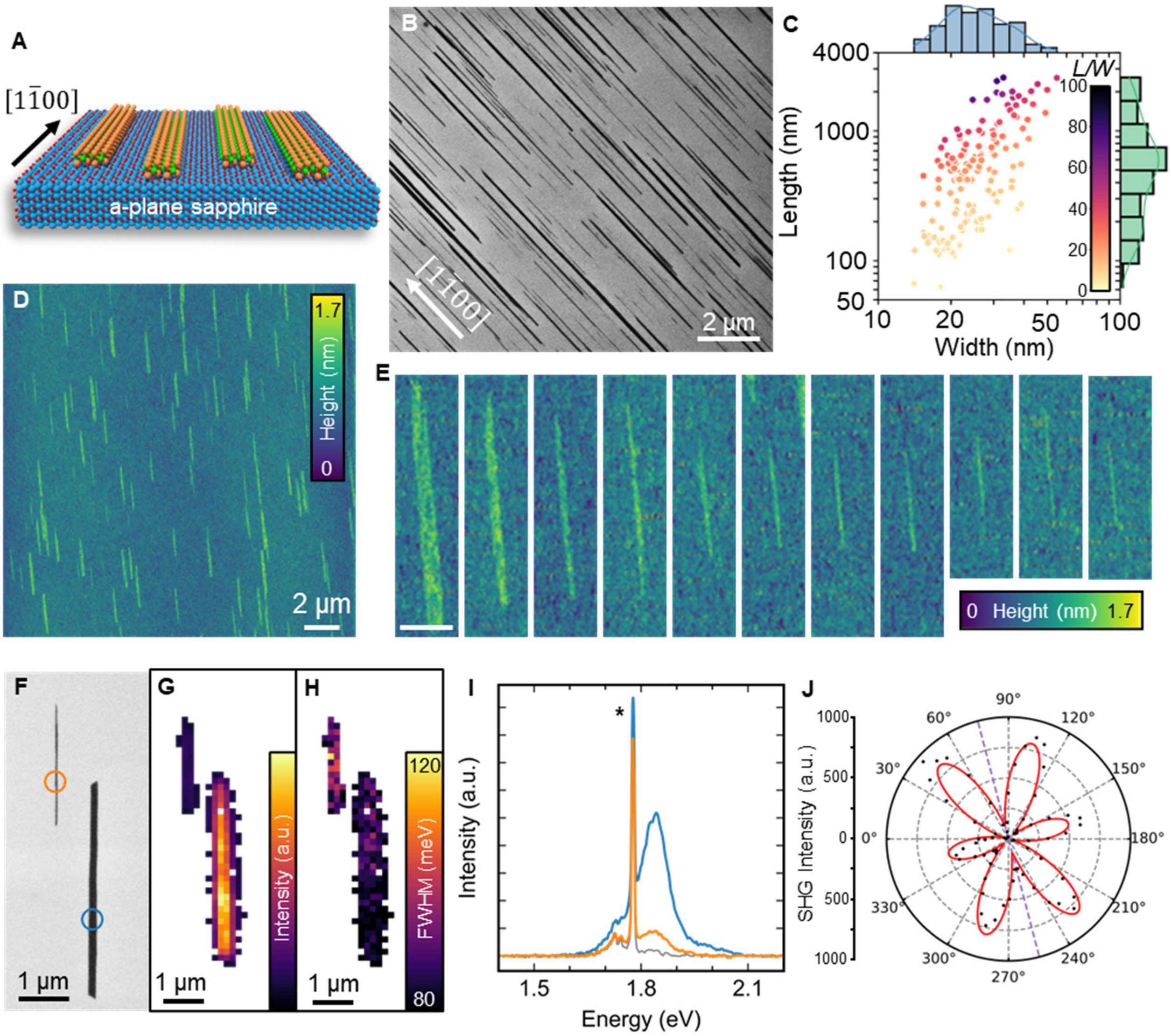

**Fig. 1: Latice-guided growth of $MoS_2$ NRs**. (**A**) Model of the aligned $MoS_2$ NRs on a-sapphire. (**B**) SEM image of the as-grown NRs. (**C**) Distribution of the width versus the length of ~150 NRs, measured by SEM. The color of the markers indicate the aspect ratio (*L/W*) of the given NR. (**D**) AFM image of the as-grown NRs. (**E**) High magnification AFM images of isolated narrow NRs, with widths down to ~24 nm. The scalebar in (E) represents 500 nm. SEM image (**F**) and corresponding PL mapping of the A-exciton intensity (**G**) and FWHM (full width at half maximum) (**H**) of two nanoribbons of widths 40 nm (left) and 130 nm (right). (**I**) PL spectra of the narrow (orange) and wide (blue) ribbons, taken at the positions marked in (F). The spectrum of bare sapphire (gray) is shown for comparison, and the signal from the sapphire is marked with an asterisk. (**J**) Polar plot of the polarization-resolved SHG signal intensity of the $MoS_2$ NRs. The black dots correspond to the experimental data while the red line is the fit. The longitudinal axis of the NR is indicated by the purple dashed line. The PL in (G-I) was collected at room temperature using an excitation wavelength of 532 nm.

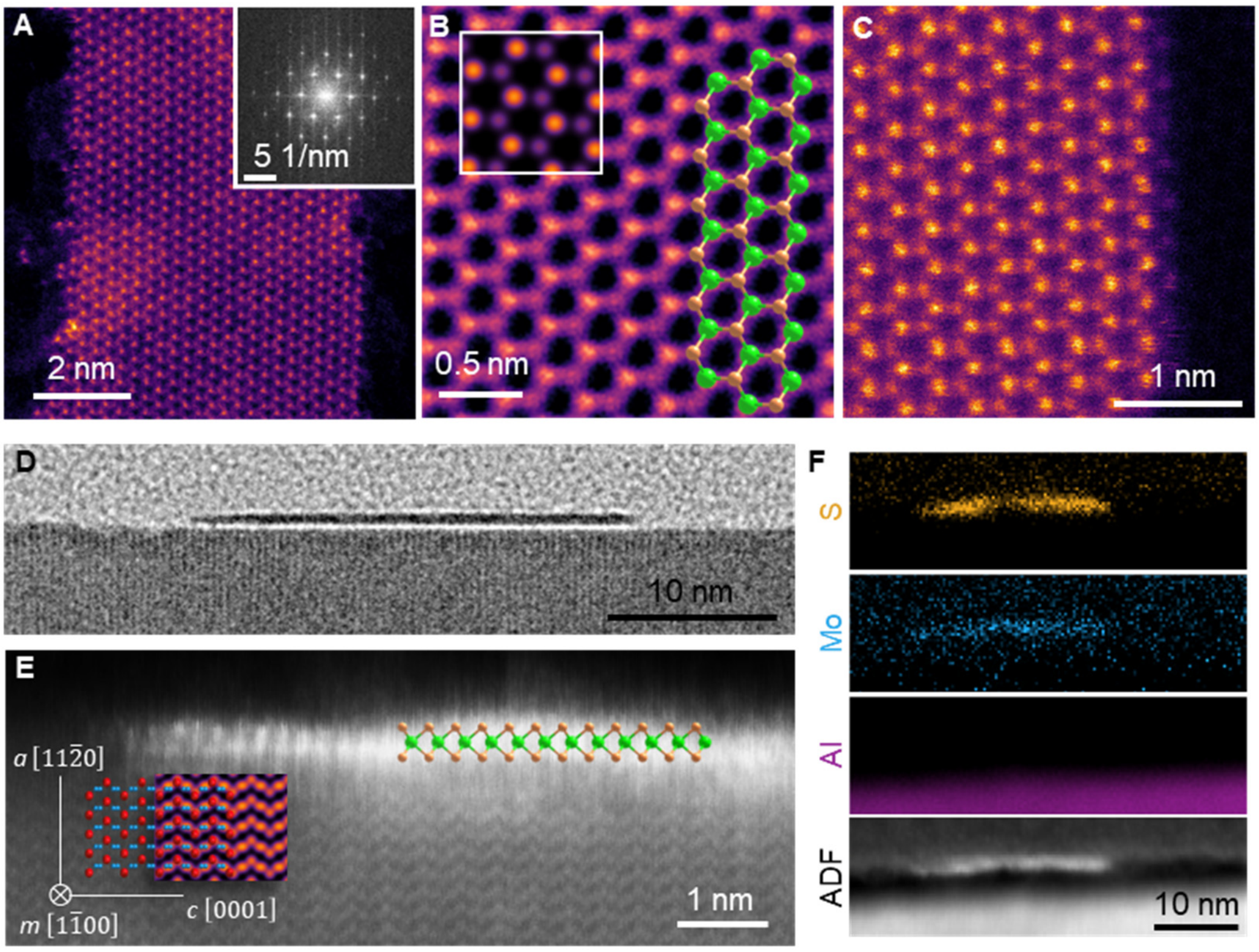

**Fig. 2: Structure and alignment of the $MoS_2$ NRs**. (**A-C**)Top-view STEM images of a 5.7 nm-wide $MoS_2$ NR supported on graphene. (**A**) Low-magnification image of the NR. The inset corresponds to the FFT. (**B**) FFT-filtered atomic resolution STEM image of the NR. The top-left inset shows a simulated STEM image of $MoS_2$. A model of the $MoS_2$ lattice is overlapped on the right side, with Mo and S atoms colored in green and orange, respectively. (**C**) High-resolution image of the Mo-S terminated zigzag edge of the NR. The row of single S atoms at the edge is clearly visible. (**D**) Cross-section TEM image of a 26 nm-wide NR on the sapphire. (**E**) STEM image of the S-terminated edge of the NR in (D). A schematic of the $MoS_2$ lattice is superimposed on the right side of the NR in (E). A schematic of the cross-section of the sapphire (m-plane, with Al and O atoms in blue and red, respectively) and a simulated STEM image are superimposed at the bottom-left side in (E). (**F**) EELS element mappings of S (158.0-203.5 eV), Mo (248.8-276.3 eV), and Al (68.8-128.3 eV), and the corresponding annular dark-field (ADF) image.

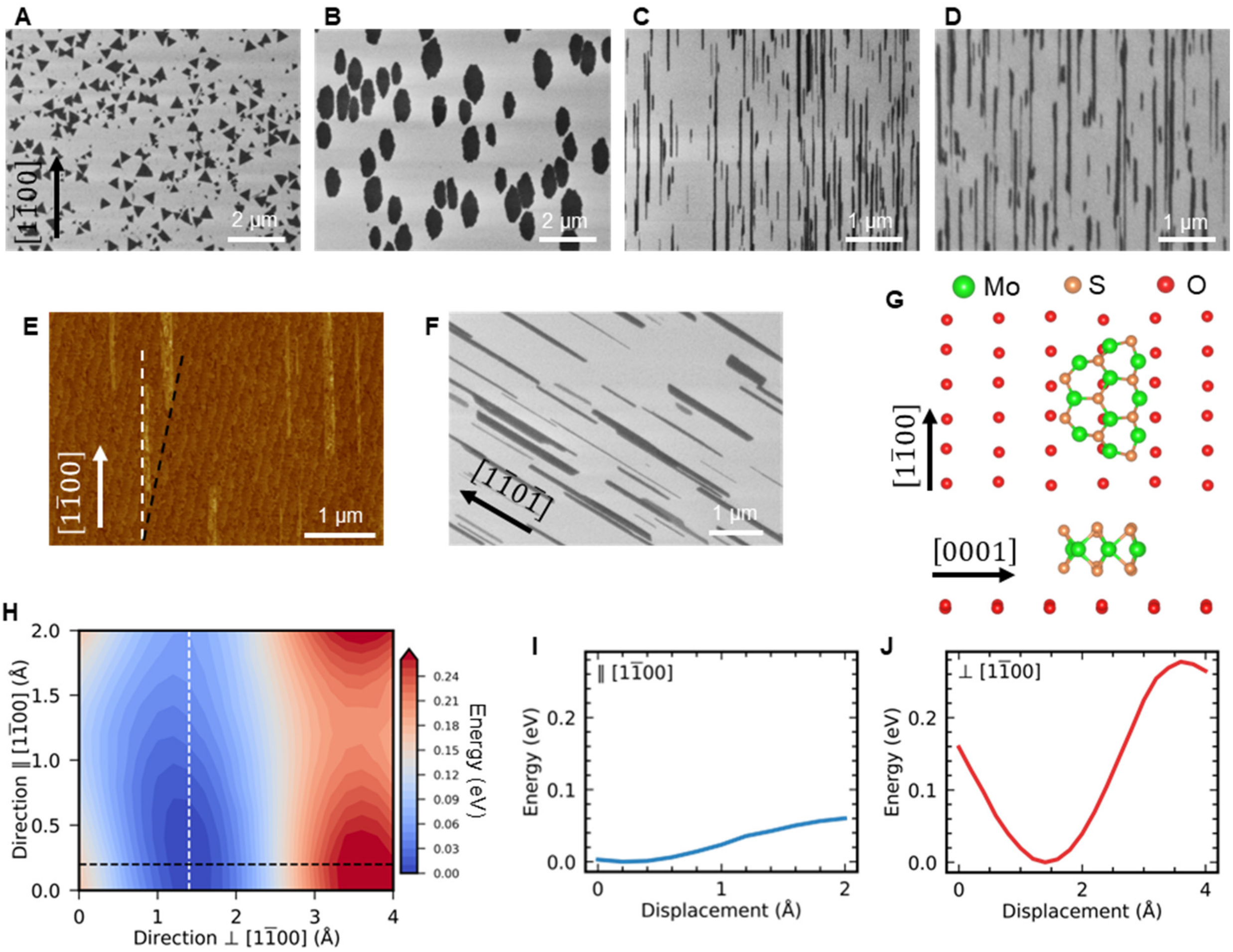

**Fig. 3: Growth mechanism of aligned $MoS_2$ NRs**. (**A-D**) Temperature dependence of the morphological structure of $MoS_2$ grown at 750 °C (A), 950 °C (B), 1100 °C (C), and 1200 °C (D). The increase in the CVD temperature produces both an alignment and a 1D constrained growth. (**E**) AFM image showing a relative alignment of the $MoS_2$ NRs (dashed while line) and the atomic steps of the sapphire surface (dashed black line) of around 13°. (**F**) $MoS_2$ NRs grown on r-plane sapphire. (**G**) Atomic model used for the DFT calculations viewed from the top $[11\bar{2}0]$ direction (upper schematic) and the side $[1\bar{1}00]$ direction (lower schematic) of the a-plane sapphire lattice. The NR-like $MoS_2$ flake is oriented with the ZZ direction parallel to $[1\bar{1}00]$. For simplicity, only the top O atoms of the sapphire are displayed. (**H**) Contour plot of the total energy of the system shown in (G) for different positions of the NR-like $MoS_2$ flake over the sapphire. (**I-J**) Energy profiles taken along the white (I) and black (J) dashed lines in (H), crossing at the most stable configuration.

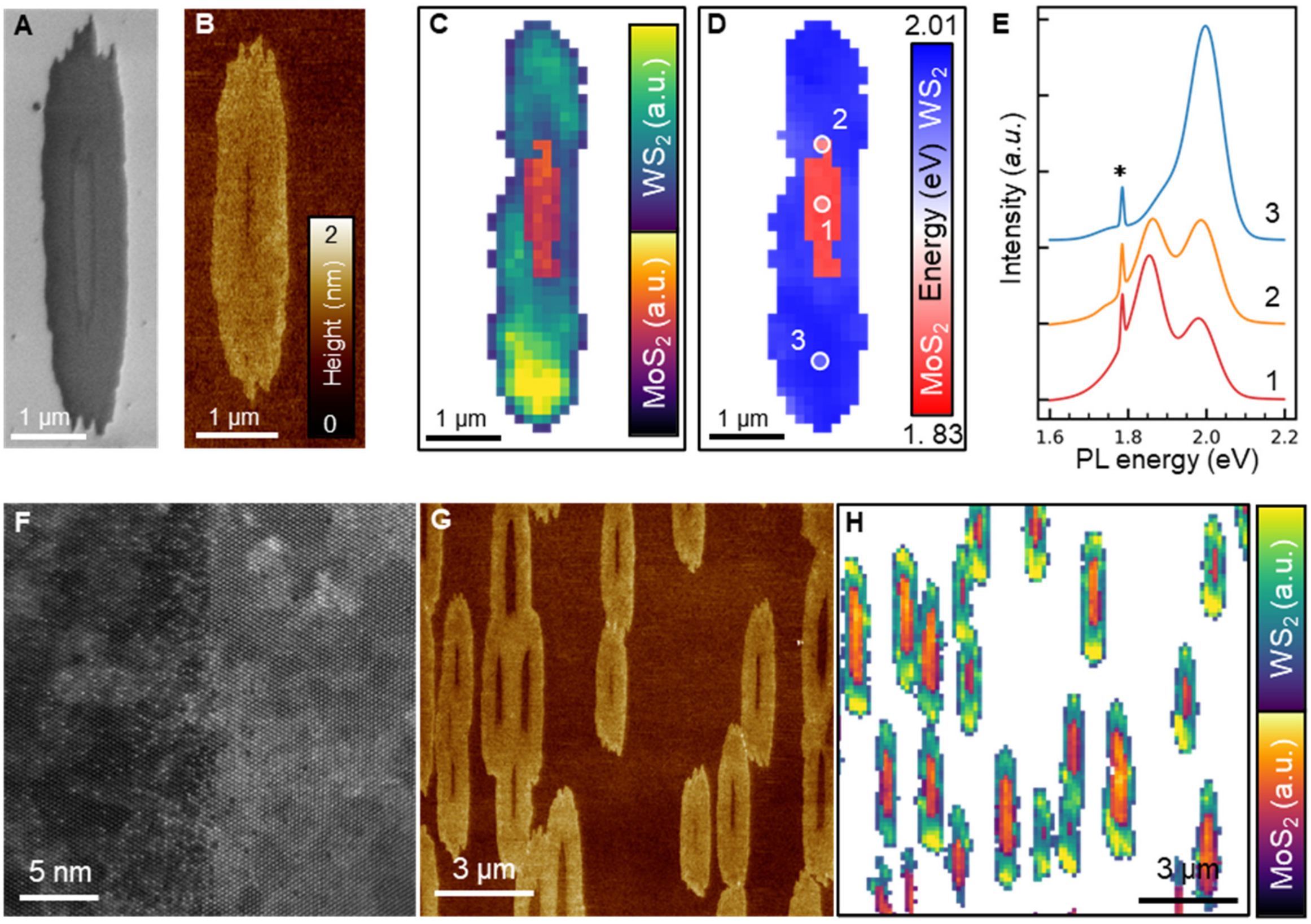

**Fig. 4: $MoS_2$-$WS_2$ hetero-nanoribbons.** (**A-B**) SEM (A) and AFM (B) images of isolated $MoS_2$-$WS_2$ lateral hNRs. (**C-D**) Mappings of the PL intensity (C) and the PL energy (D) of a $MoS_2$-$WS_2$ hNR. (**E**) PL spectra of the hNR collected at the points marked in (D). The asterisk in (E) indicates the signal from the sapphire substrate. (**F**) STEM image of the $MoS_2$-$WS_2$ interface, with the $MoS_2$ and the $WS_2$ on the left and right sides, respectively. (**G-H**) Wide area AFM image (G) and PL intensity mapping (H) of hNRs. Note that the SEM, AFM, and PL areas for both the isolated hNRs and the wide-area regions correspond to representative areas, but are not identical. The PL data was collected at room temperature using an excitation wavelength of 532 nm.

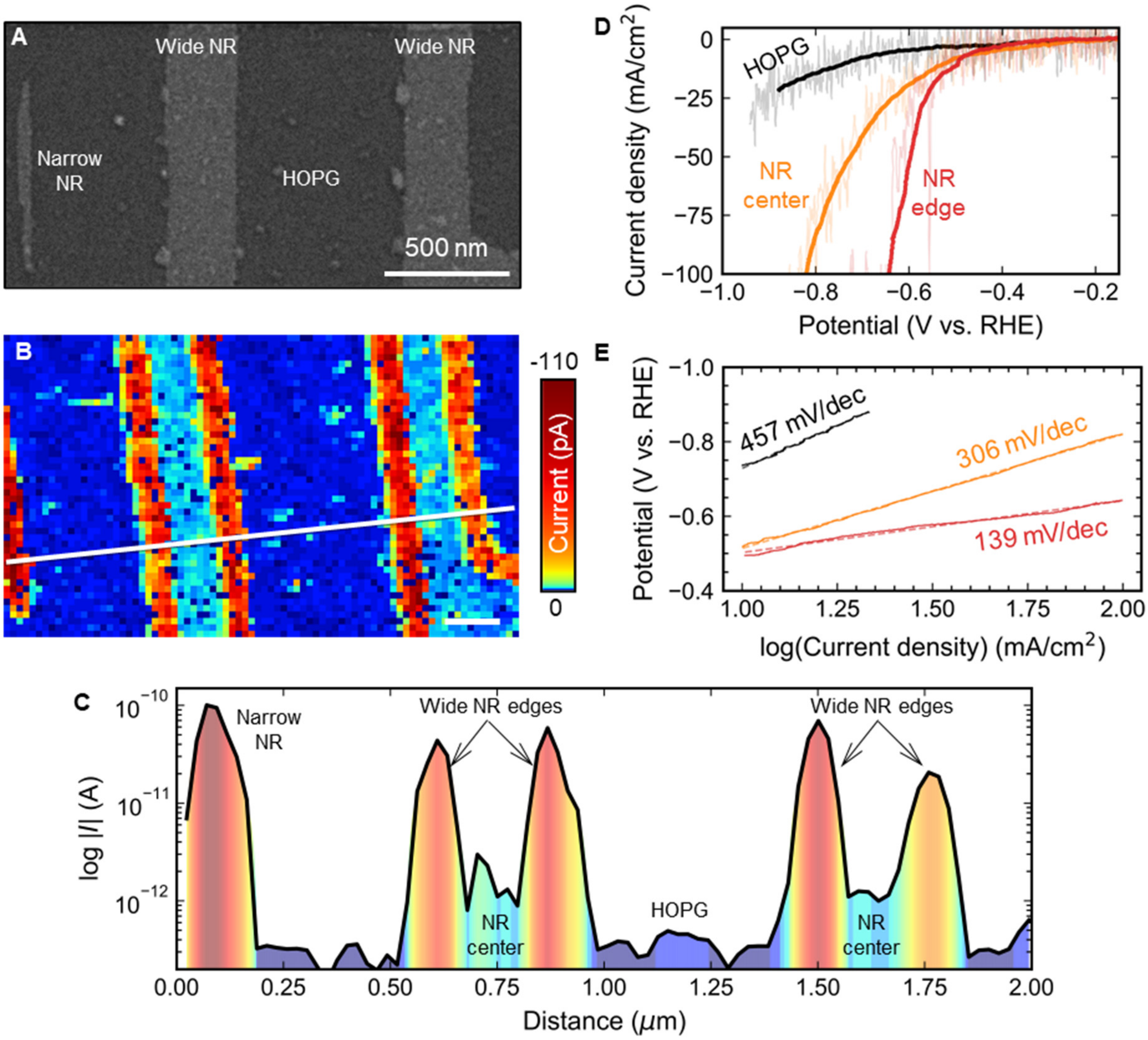

**Fig. 5: Catalytic activity of the $MoS_2$ NRs.** (**A-B**) SEM image (A) and the corresponding HER current (-0.95V vs RHE) mapping (B) of a narrow NR and two wide NRs supported on HOPG. The scale bar in (B) represents 200 nm. (**C**) Electrochemical current along the white line indicated in (B). The labels indicate the nature of the corresponding regions. The shown profile is averaged from 5 parallel profiles to decrease the noise. (**D-E**) Overpotential (D) and Tafel slope (E) for HOPG (black), and for the center (orange) and edge (red) of a wide NR. The overpotential graph (D) displays both the raw data (faded colors) and the filtered signals (solid colors) to demonstrate the effect of data processing and highlight underlying trends.

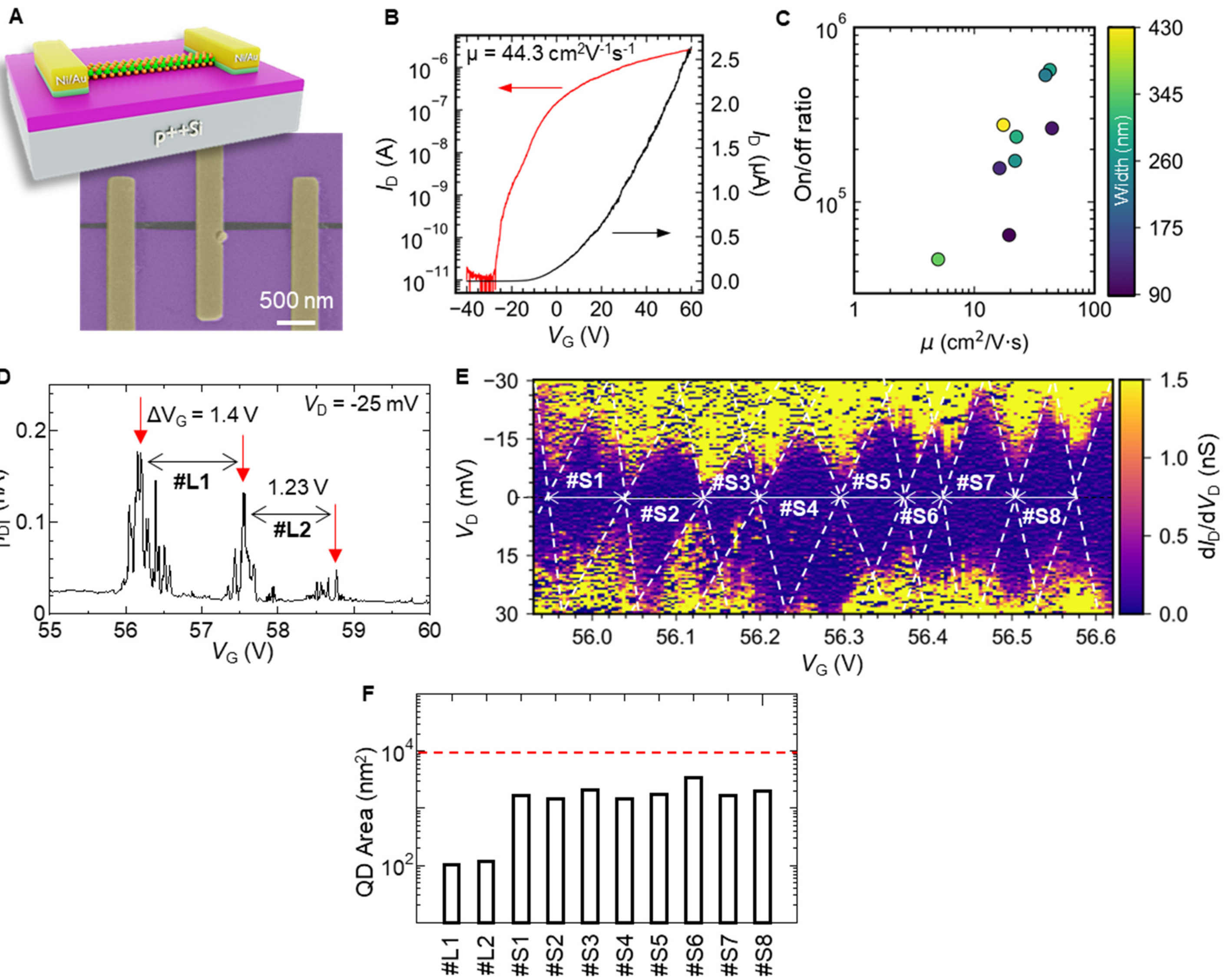

**Fig. 6: Electrical performance of the $MoS_2$ NRs.** (**A**) Schematic (top) and false-colored SEM image of an actual NR device (bottom). (**B**) Room temperature transfer characteristics of a FET with a NR channel of width 110 nm. (**C**) Mobility vs on/off ratio of several NR devices, collected at room temperature. The width of the NRs is indicated by the color of the markers. (**D**) Long-period Coulomb oscillation peaks of a device with a NR channel of width 30 nm measured at 4.2 K. The red arrows indicate the periodicities. (**E**) Edge-highlighted short-period Coulomb diamond features of the 30-nm wide NR device. (**F**) Comparison of whole NR area (9000 $nm^2$, indicated by the red dashed line) and calculated QD sizes from (D) and (E).

# Supplementary Materials

# Lattice-guided growth of dense arrays of aligned transition metal dichalcogenide nanoribbons with high catalytic reactivity

Z. Ma,[1] P. Solís-Fernández,[2*] K. Hirata,[3] Y.-C. Lin,[4,5] K. Shinokita,[6] M. Maruyama,[7] K. Honda,[3] T. Kato,[8] A. Uchida,[2] H. Ogura,[8] T. Otsuka,[8,9,10,11,12] M. Hara,[13,14] K. Matsuda,[6] K. Suenaga,[5] S. Okada,[7] T. Kato,[8,10] Y. Takahashi,[3,15] & H. Ago[1,2,16*]

[1] Interdisciplinary Graduate School of Engineering Sciences, Kyushu University, Japan.

[2] Faculty of Engineering Sciences, Kyushu University, Fukuoka 816-8580, Japan

[3] Department of Electronics, Graduate School of Engineering, Nagoya University, Japan.

[4] Nanomaterials Research Institute, National Institute of Advanced Industrial Science and Technology (AIST), Tsukuba 305-8565, Japan.

[5] The Institute of Scientific and Industrial Research, Osaka University, Japan.

[6] Institute of Advanced Energy, Kyoto University, Kyoto, 611-0011, Japan.

[7] Department of Physics, Graduate School of Pure and Applied Sciences, University of Tsukuba, Tsukuba 305-8571, Japan.

[8] Graduate School of Engineering, Tohoku University, Sendai 980-8579, Japan.

[9] Research Institute of Electrical Communication, Tohoku University, Sendai 980-8577, Japan.

[10] Advanced Institute for Materials Research (AIMR), Tohoku University, Japan.

[11] Center for Science and Innovation in Spintronics, Tohoku University, Japan

[12] Center for Emergent Matter Science, RIKEN, Saitama 351-0198, Japan

[13] Faculty of Advanced Science and Technology, Kumamoto University, Japan.

[14] Institute of Industrial Nanomaterials, Kumamoto University, Japan.

[15] WPI Nano Life Science Institute (WPI-NanoLSI), Kanazawa University, Japan.

[16] Center for Semiconductor and Device Ecosystem (CSeDE), Kyushu University, Japan.

*Corresponding authors: Pablo Solís-Fernández (solisfernandez.pablo.kyudai@gmail.com), Hiroki Ago (ago.hiroki.974@m.kyushu-u.ac.jp)

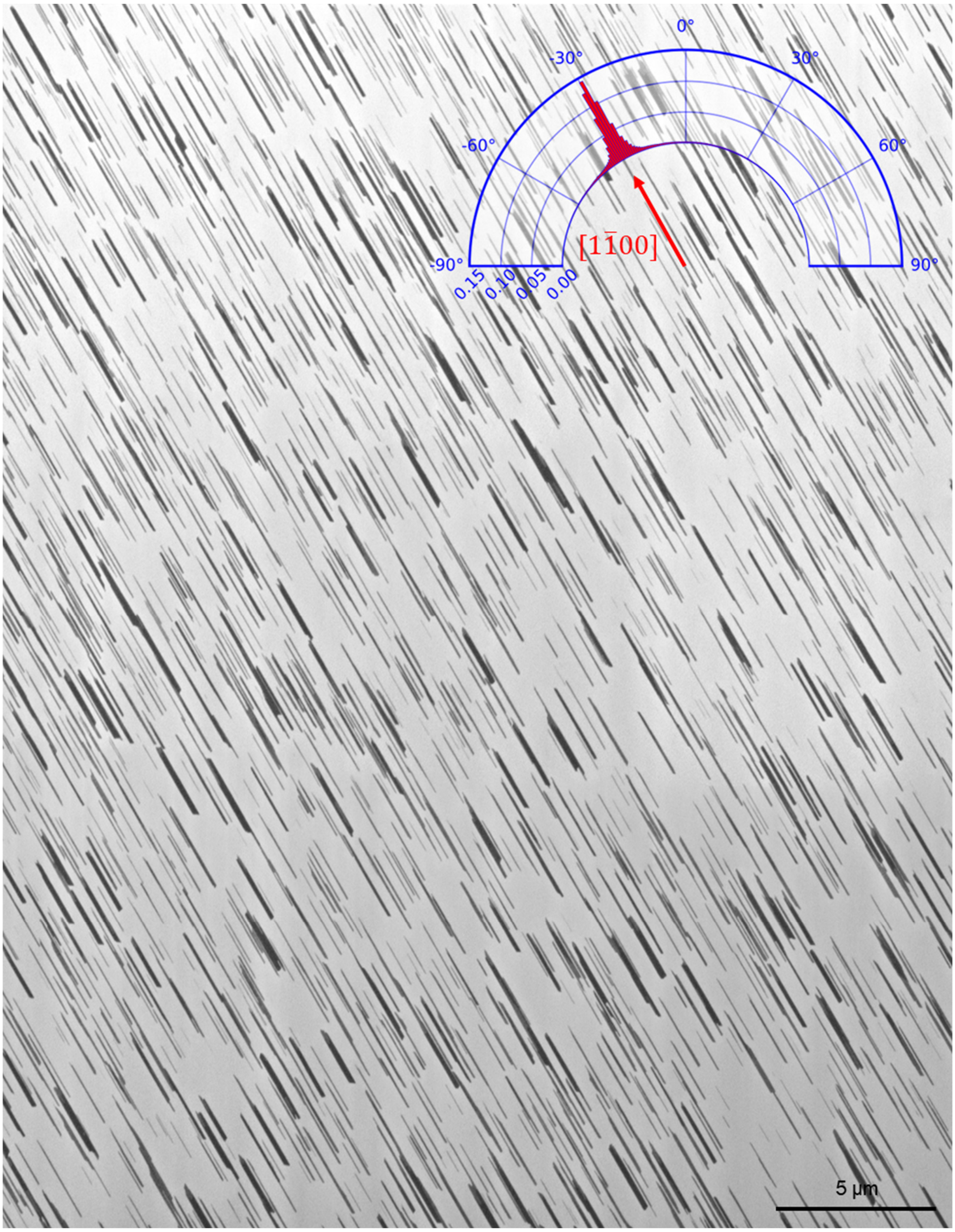

**Fig. S1**. Stitched SEM image of $MoS_2$ nanoribbons on a-plane sapphire. The inset shows the histogram in polar coordinates of the alignment of the ribbons in the area, evidencing a narrow distribution parallel to the [*1$\bar{1}$00*] direction of sapphire indicated by the red arrow.

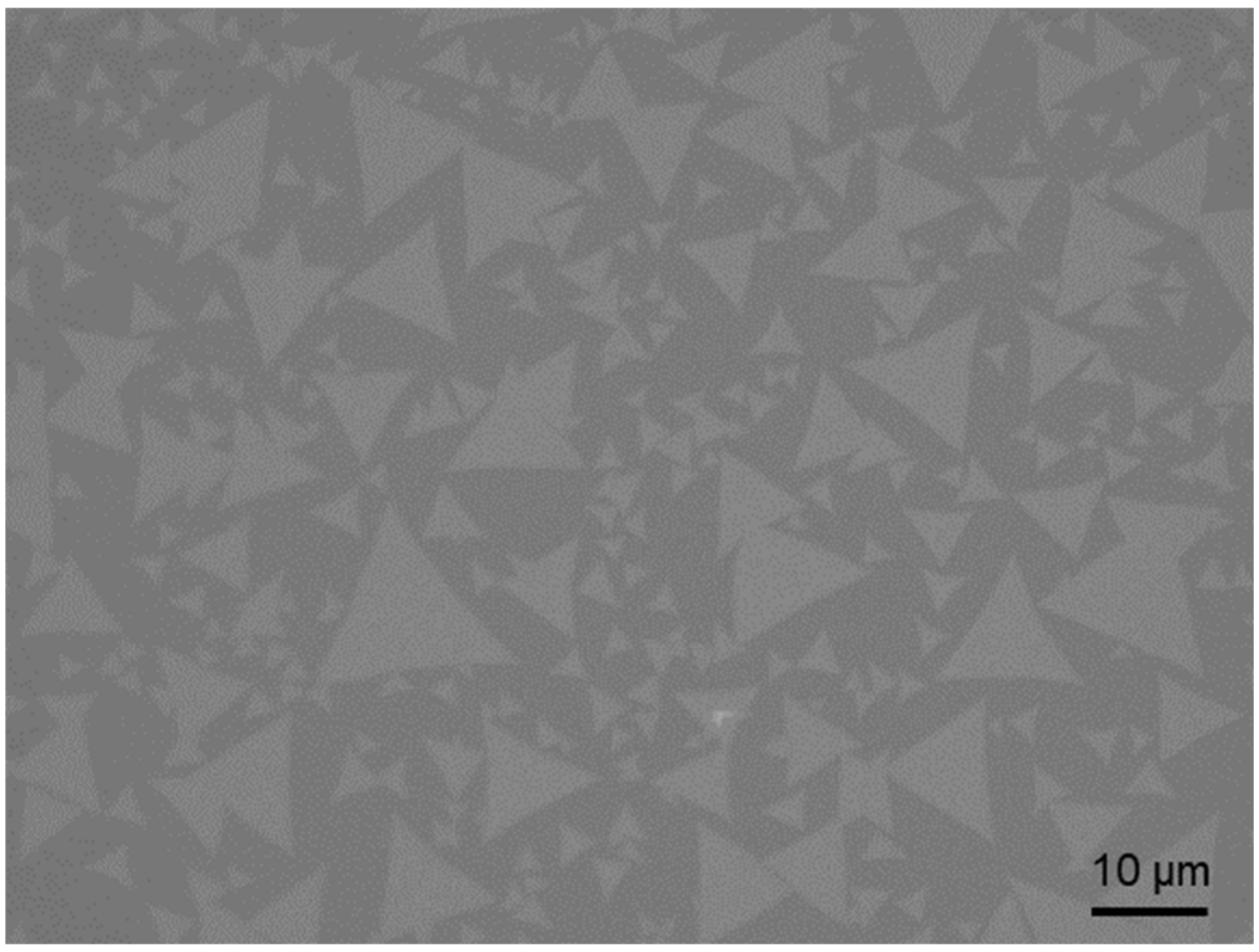

**Fig. S2**. Optical microscope image of $MoS_2$ grown on c-plane sapphire using the same CVD growth condition (1100 °C) as for the sample shown in Fig. 1B. Triangular flakes were obtained on the c-plane sapphire instead of the oriented NRs obtained on a-sapphire.

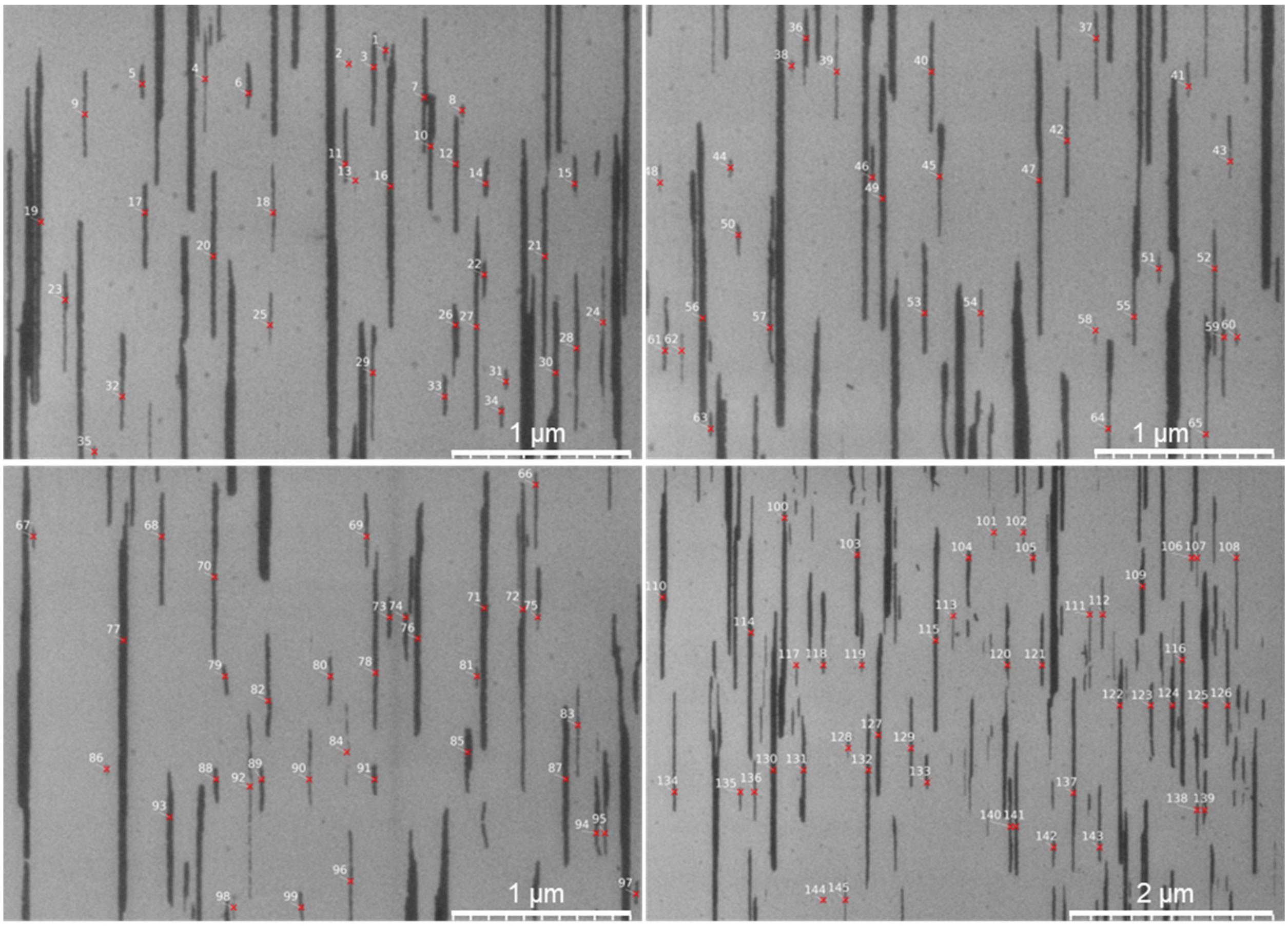

**Fig. S3**. SEM image showing the $MoS_2$ NRs from which statistics used in Fig. 1C were collected.

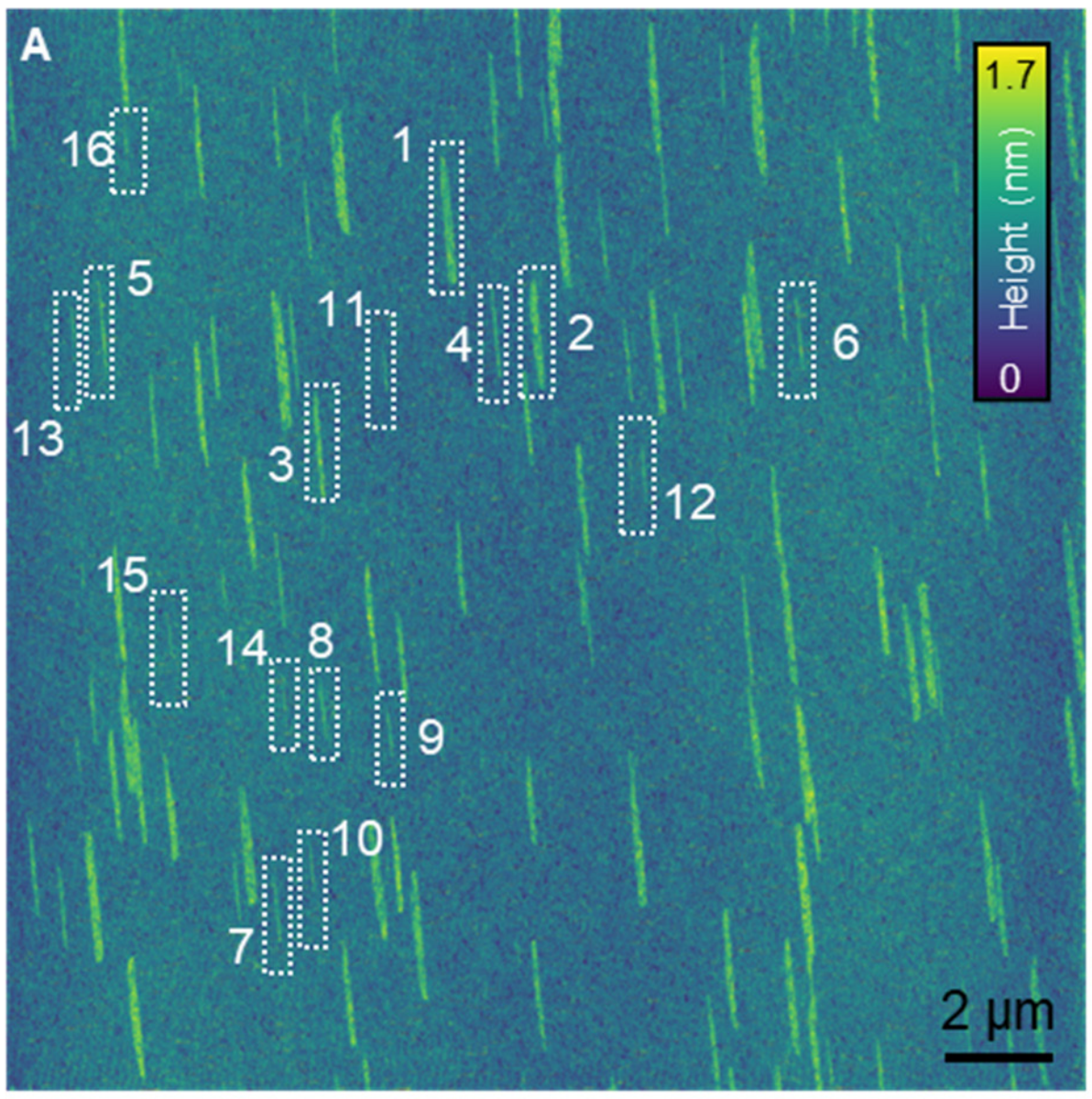

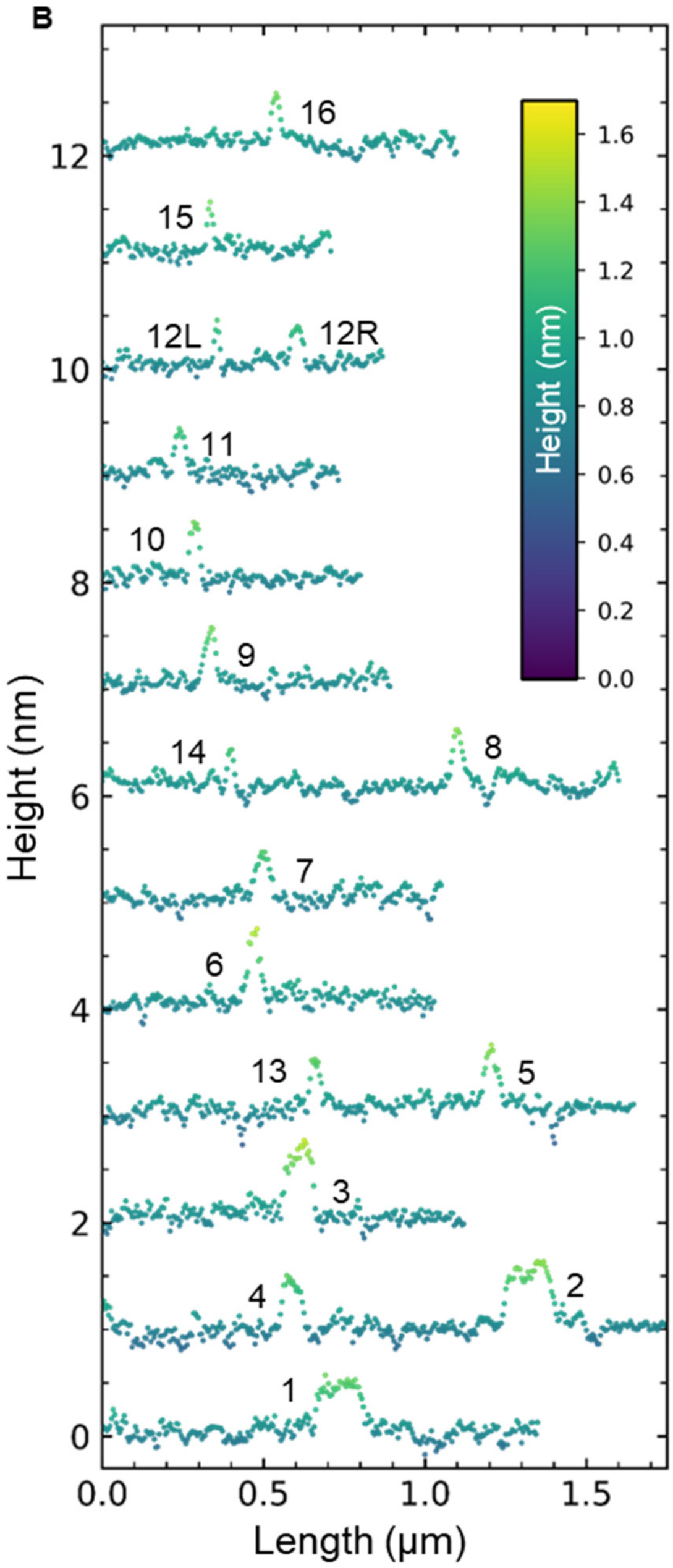

**C**

| Ribbon # | Width (nm) | Height (Å) |
|---|---|---|
| 1 | 147.1 | 4.1 |
| 2 | 151.5 | 5.2 |
| 3 | 92.9 | 5.8 |
| 4 | 68.4 | 4.6 |
| 5 | 48.8 | 5.0 |
| 6 | 58.8 | 6.1 |
| 7 | 58.7 | 4.0 |
| 8 | 39.2 | 4.8 |
| 9 | 44.1 | 4.9 |
| 10 | 34.1 | 4.9 |
| 11 | 39.0 | 3.9 |
| 12 (left) | 24.4 | 4.1 |
| 12 (right) | 48.8 | 3.4 |
| 13 | 34.2 | 4.0 |
| 14 | 24.5 | 3.3 |
| 15 | 24.6 | 4.0 |
| 16 | 29.4 | 4.2 |

**Fig. S4**. Size of the $MoS_2$ nanoribbons by AFM. (**A**) Large-area AFM image of the NRs shown in Fig. 1D, E. The selected NRs are highlighted by white dotted rectangles. (**B**) Height profiles across each of the NR highlighted in (A). The profiles were vertically shifted for clarity, and the color indicates the actual height. (**C**) Measured values of width and height of each of the NRs of (A) and (B). The average height of the NRs is ~4.5 Å, indicating their single-layer nature.

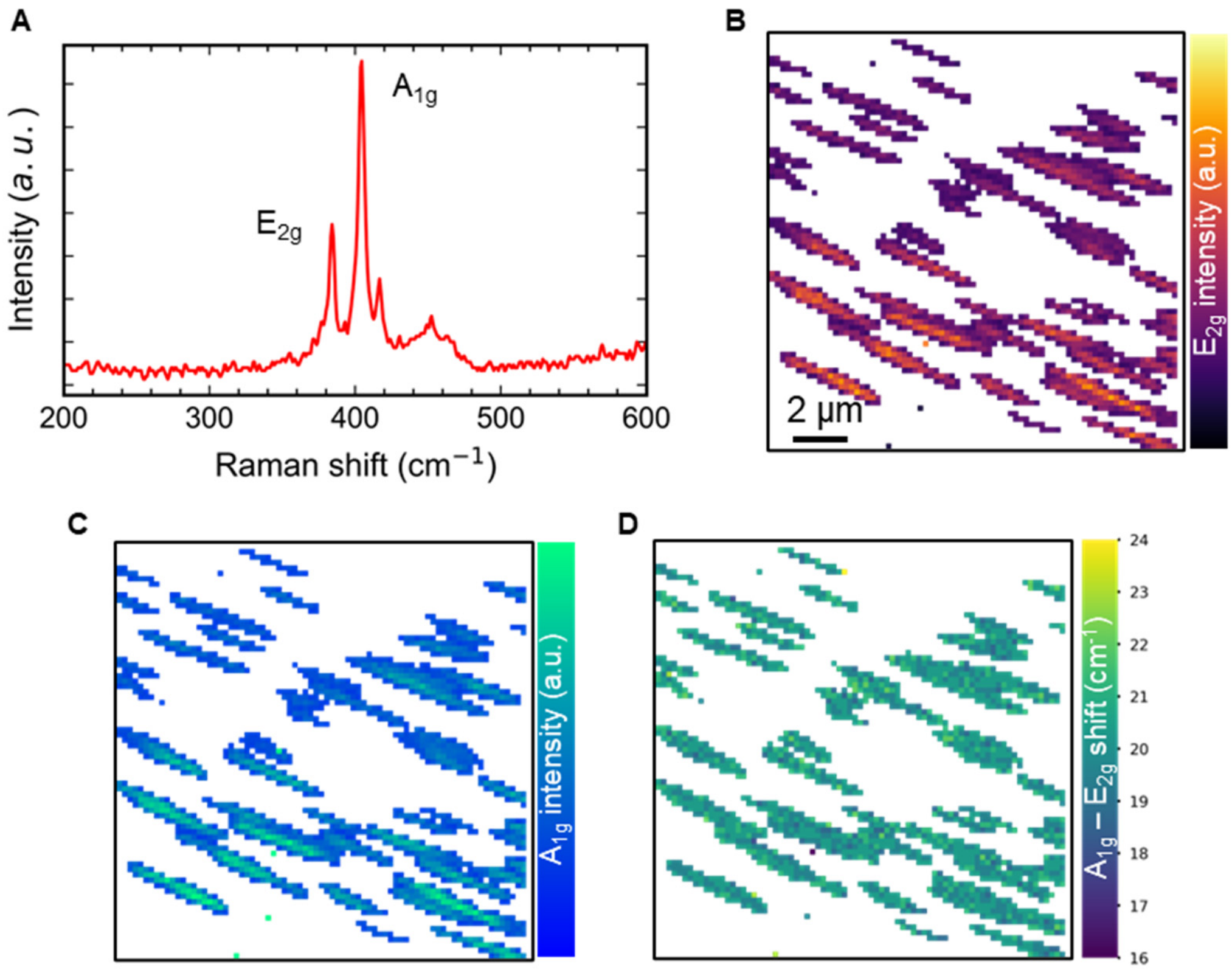

**Fig. S5**. Raman characterization of the $MoS_2$ nanoribbons. (**A**) Raman spectrum of a $MoS_2$ NR. (**B-D**) Raman mappings of the $E_{2g}$ (**B**) and $A_{1g}$ (**C**) intensity, and of the ($A_{1g}$ – $E_{2g}$) shift difference (**D**) of $MoS_2$ NRs on a-sapphire. The uniform peak distance in the Raman spectrum (~20 $cm^{-1}$) in (D) is an indicator of the single layer nature of the NRs.

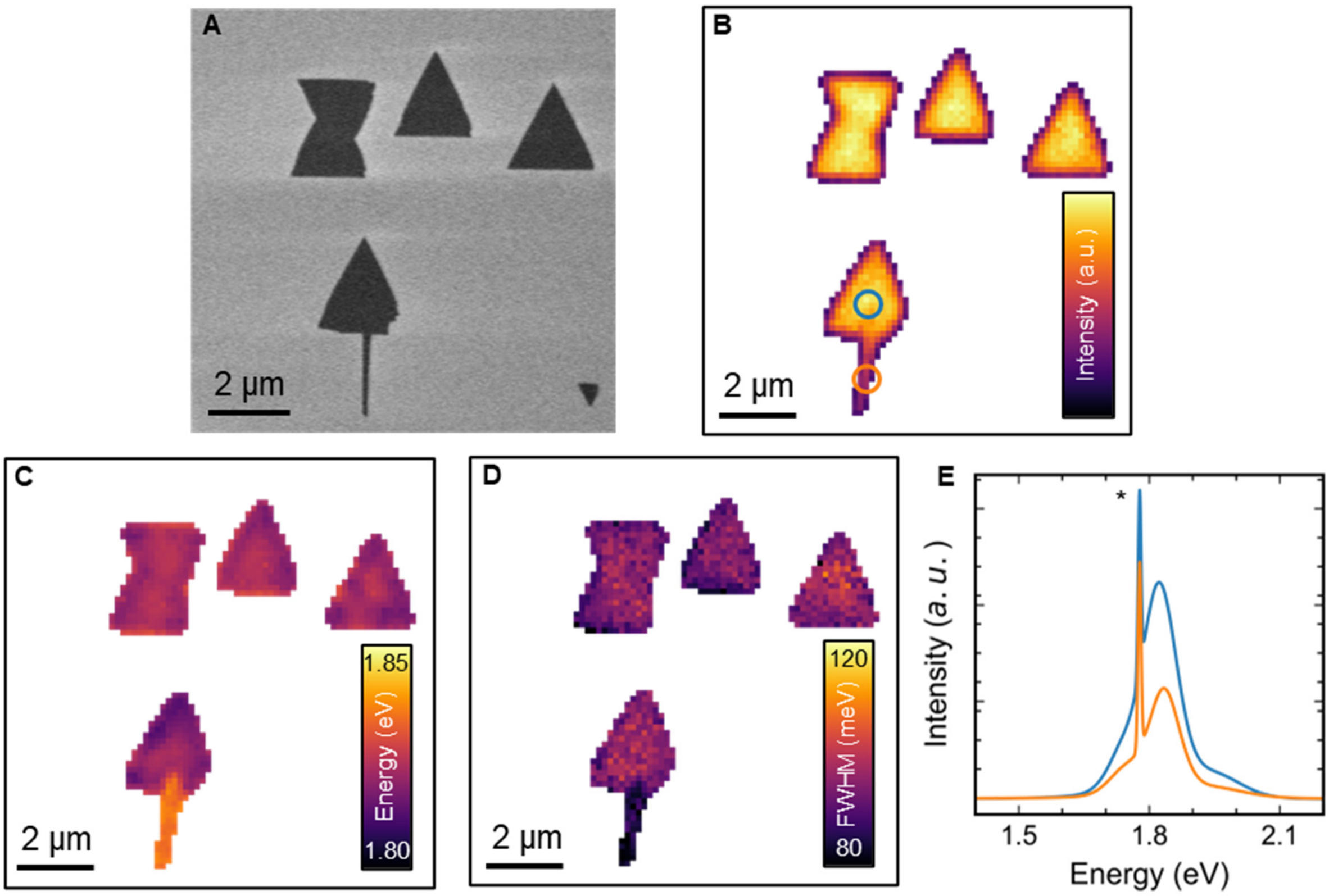

**Fig. S6**. Characterization of $MoS_2$ flakes. (**A-D**) SEM image (**A**) and the corresponding PL mappings for the intensity (**B**), energy (**C**), and full-width at half maximum (FWHM) (**D**) of the A-exciton peak (~1.8 eV) of $MoS_2$ in an area containing triangular flakes and a NR. (**E**) PL spectra collected at the positions marked in (B), corresponding to a triangle (blue) and a NR (orange). The asterisk in (E) indicates the signal from the sapphire substrate.

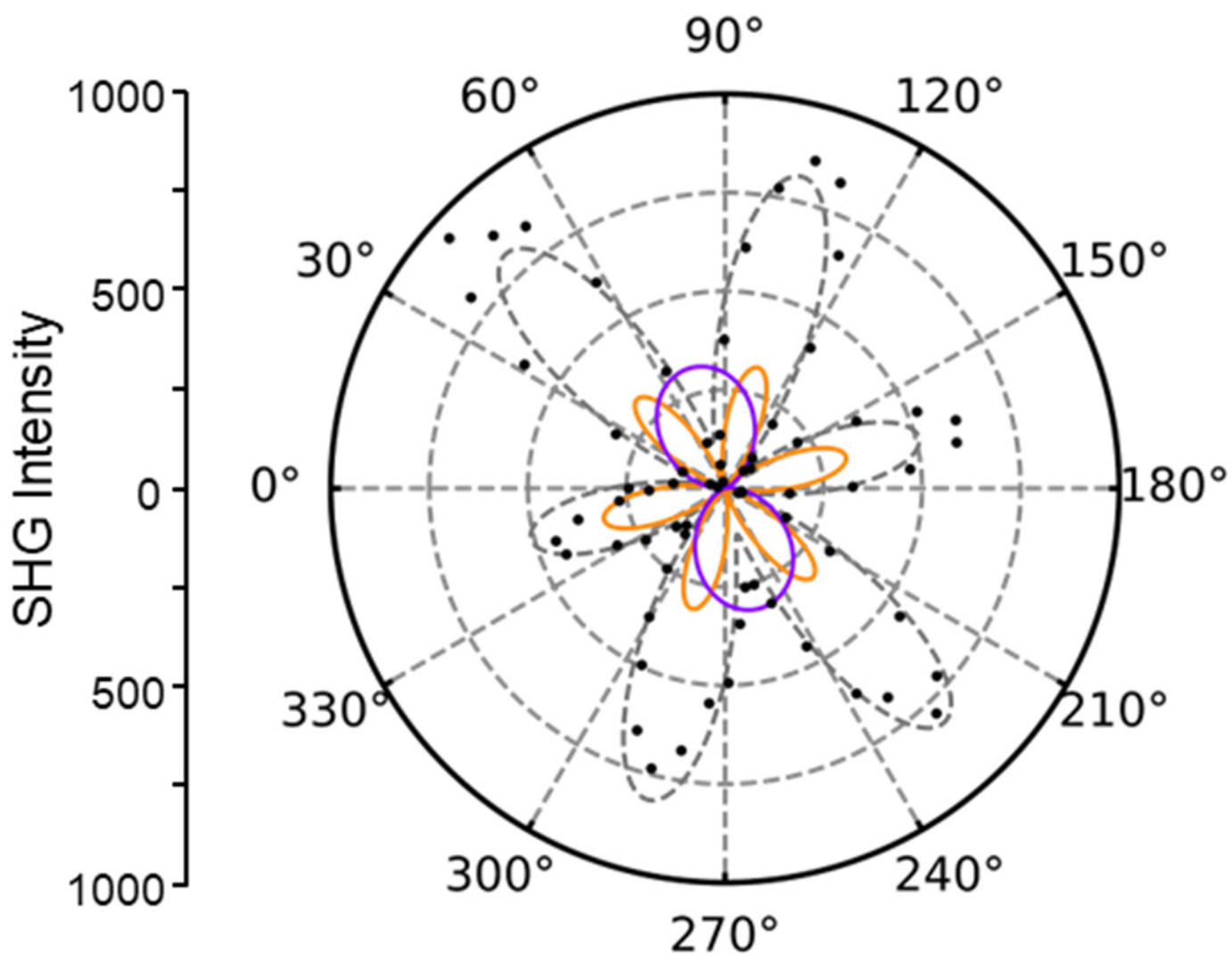

**Fig. S7**. Polar plot of the polarization-resolved SHG signal intensity with the individual $\cos^2$ components used in the fitting, with 6-fold (orange) and 2-fold (purple) symmetries. The fit result is indicated by the dashed gray line, and the black dots represent the original data. For clarity, the individual components have undergone baseline correction to shift the minimum intensity to zero.

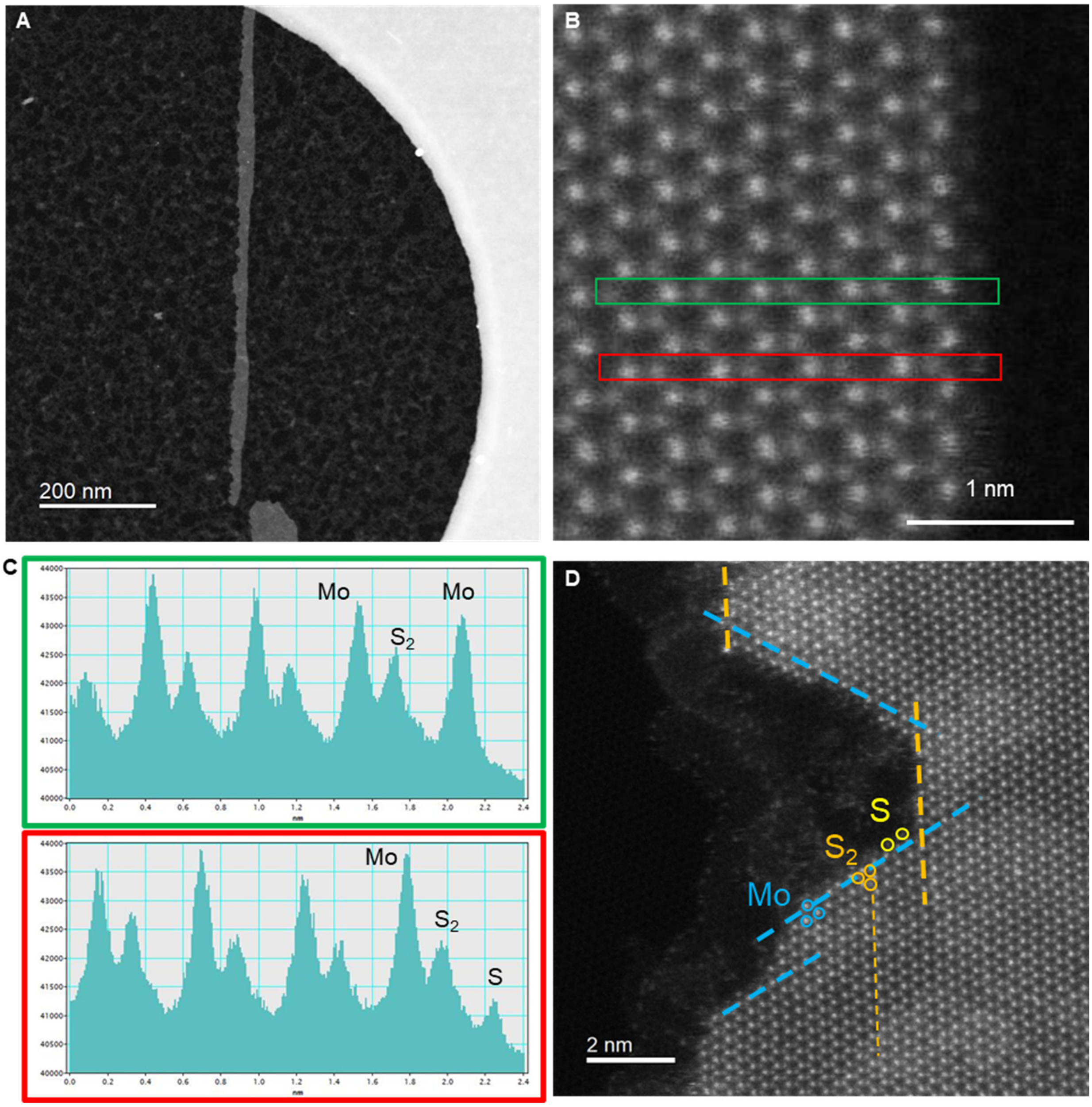

**Fig. S8**. Side-dependent edge structure of the $MoS_2$ nanoribbons. (**A**) General STEM image of a $MoS_2$ NR supported on SLG, showing the difference between the ZZ-$S_2$ edge (left) and the ZZ-Mo edge (right). (**B**) Atomic scale resolution STEM image of the ZZ-Mo edge. (**C**) Intensity profiles collected along the indicated areas in (B), showing the termination with a single S atom of the Mo atoms of the edge (ZZ-Mo(S)). (**D**) High resolution STEM image of the ZZ-$S_2$ edge. The edge retains a global direction parallel to the ZZ direction of the $MoS_2$, but it is composed by a mixture of ZZ-$S_2$ (indicated by orange dashed lines) and ZZ-Mo(S) (indicated by blue dashed lines) sections that form angles of 60°. Two of the single S atoms in the ZZ-Mo(S) edge are highlighted in yellow.

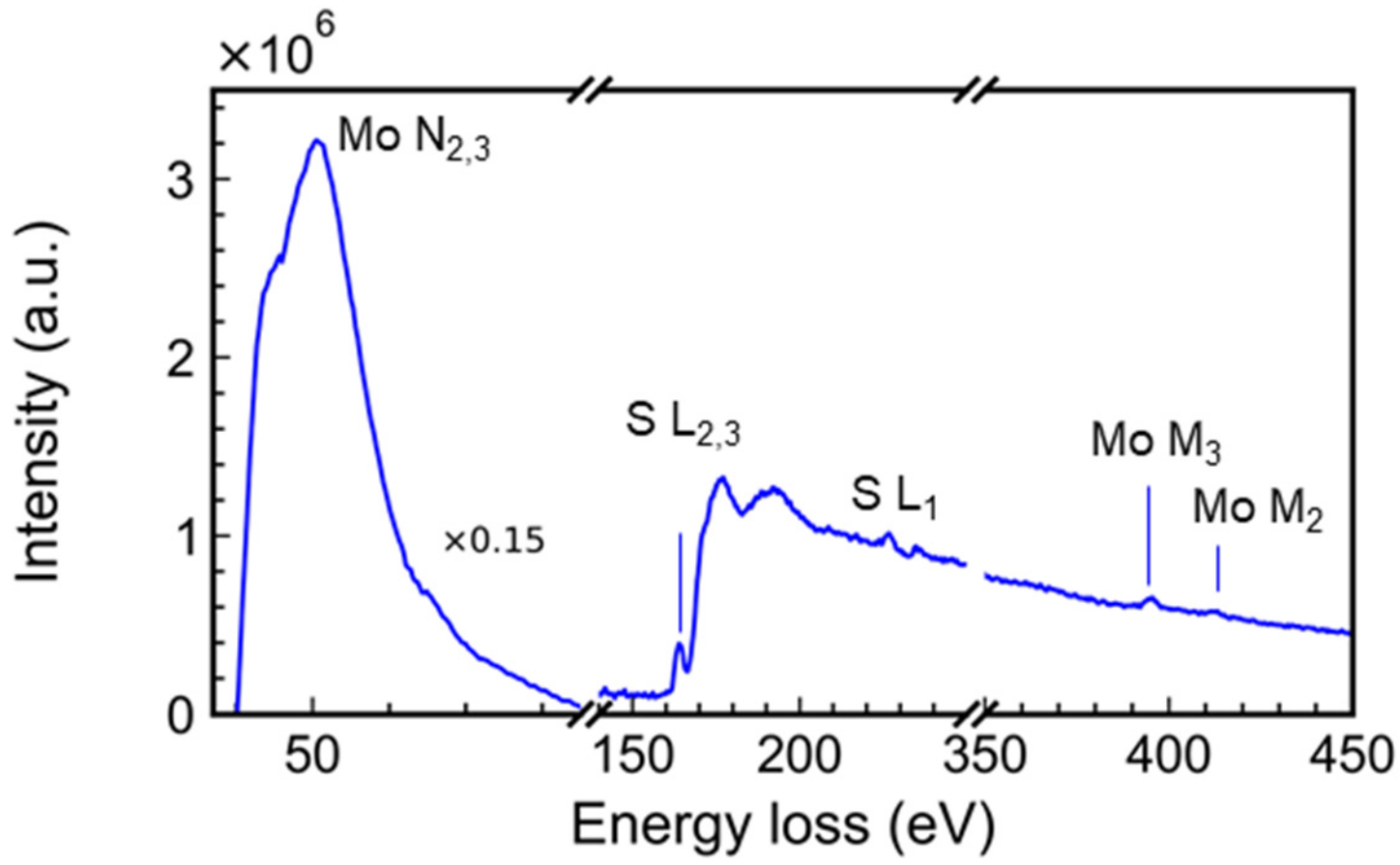

**Fig. S9**. EELS spectrum of a $MoS_2$ NR, showing the S and Mo edges.

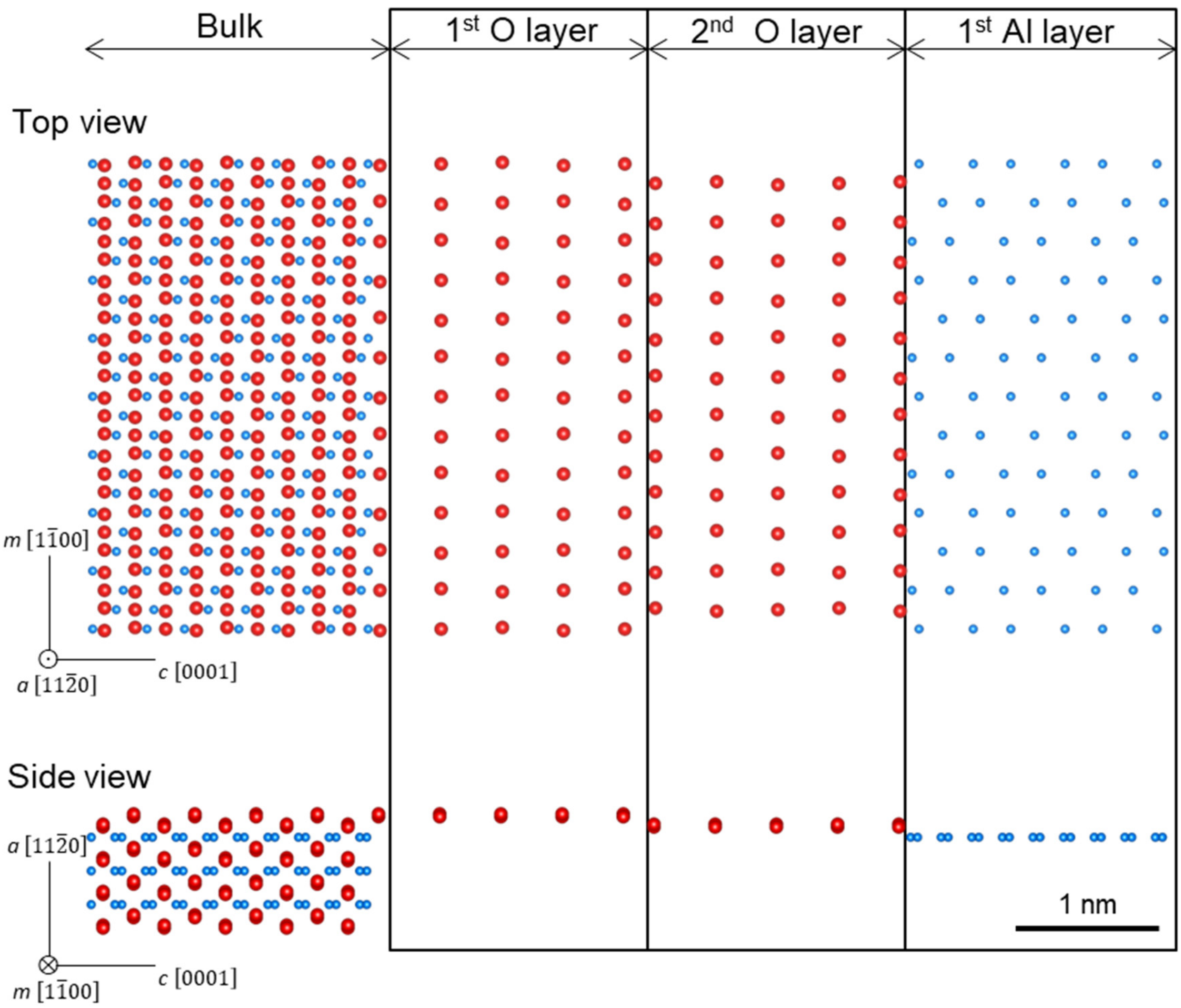

**Fig. S10**. Crystal structure of the a-plane sapphire ($11\bar{2}0$) viewed from the top and from the side (m-plane direction [$1\bar{1}00$]). The figure includes the bulk crystal structure (left side), and that of the top three atomic layers of the surface, corresponding to layers of O, O and Al atoms, respectively. Red and blue atoms represent O and Al, respectively. Linear arrays of O atoms are clearly seen along the [$1\bar{1}00$] direction.

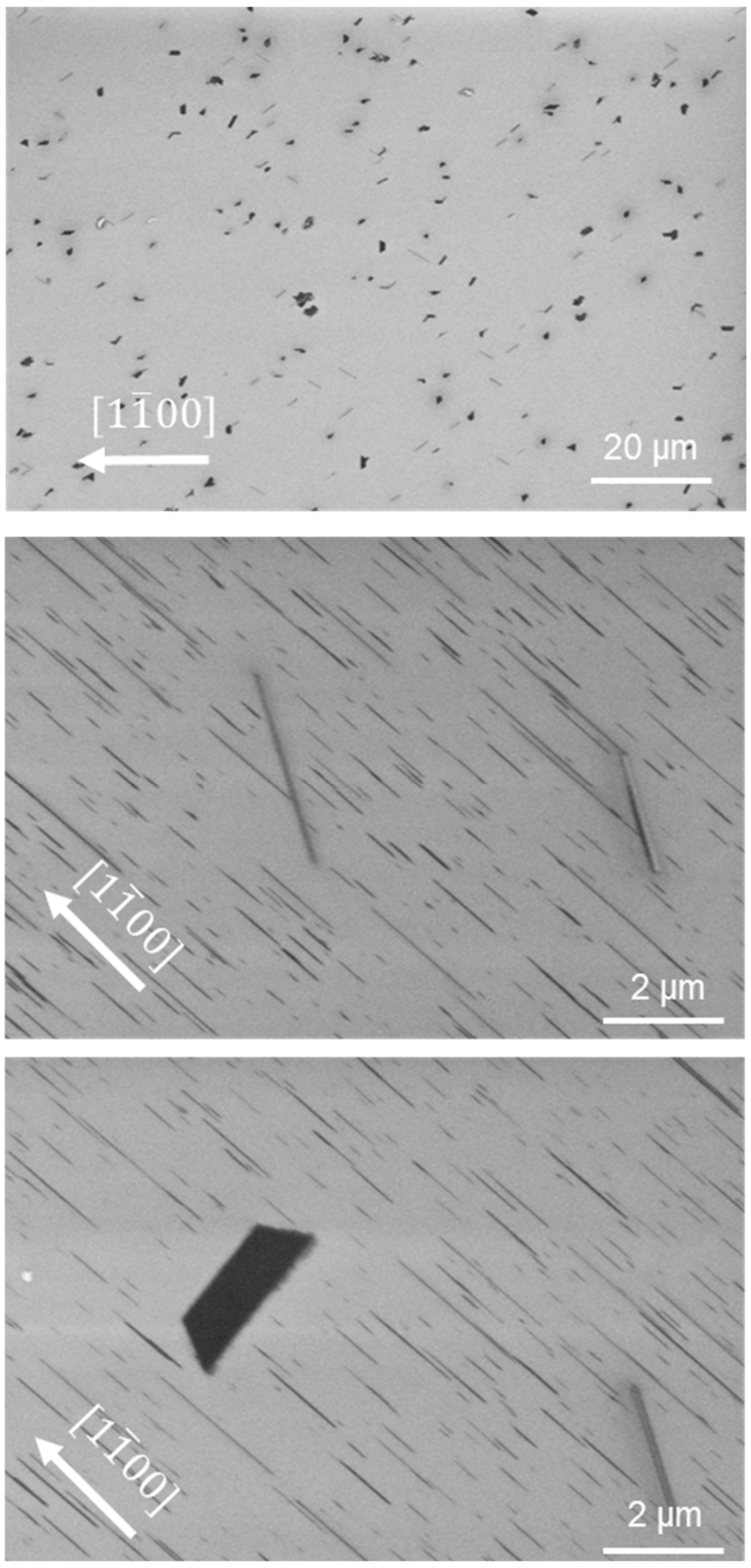

**Fig. S11**. SEM images of different regions of the a-plane sapphire after the $MoS_2$ growth at 1200 °C. The images show the presence of byproducts on the surface, and the misalignment of some of the NRs.

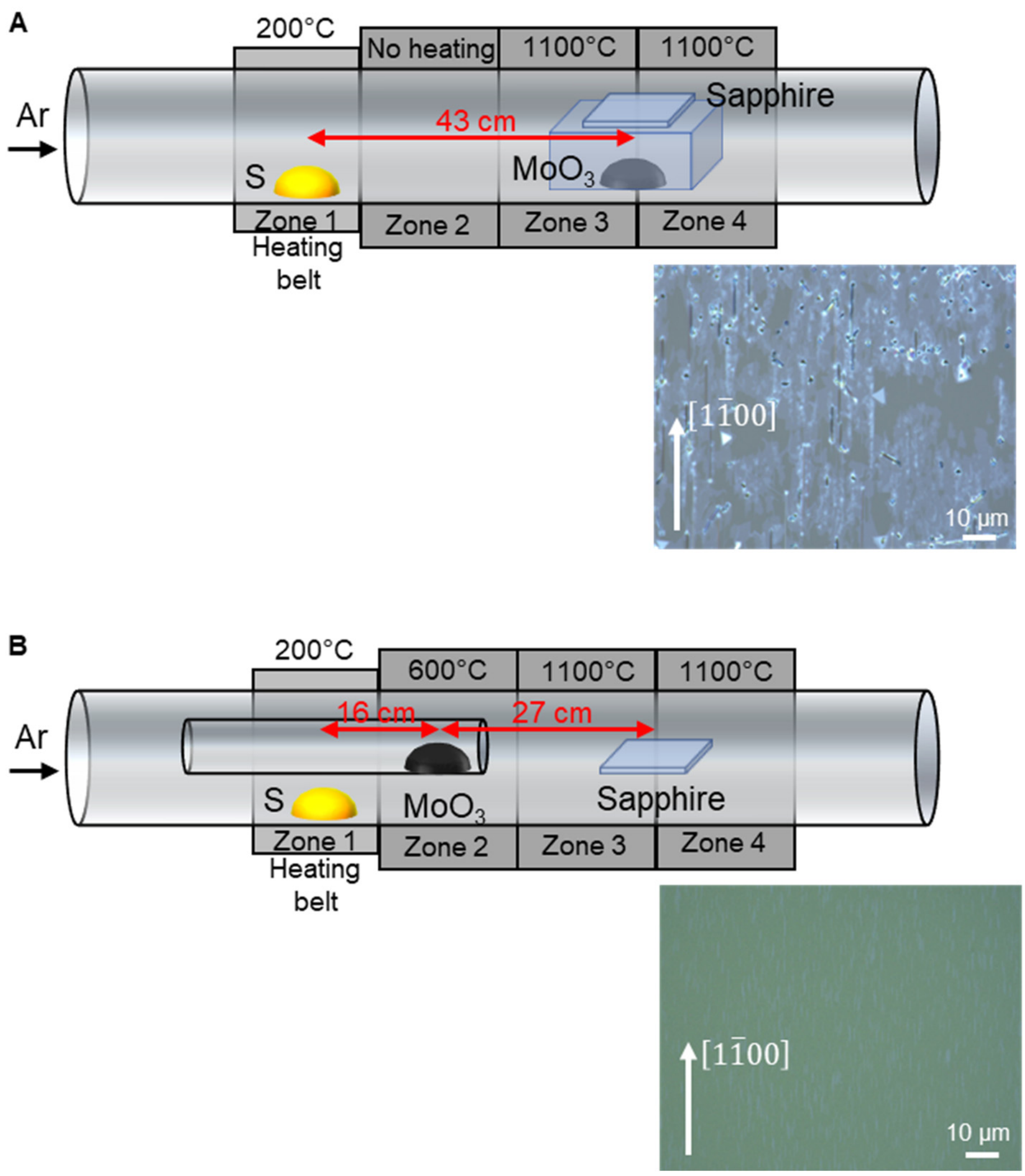

**Fig. S12**. CVD setup for the growth of nanoribbons. (**A-B**) Comparison of the face-down (**A**) and downstream (**B**) CVD setups, along with representative optical images of the respective growth at 1100 °C. The face-down setup (A) provides large amounts of precursors when the CVD temperature is increased, making difficult to control the growth of NRs and producing the growth of thick materials. Placing the substrate downstream of the precursors (B) limits the amount of $MoO_x$ and allows for the growth of narrow NRs. Moreover, the $MoO_3$ was located inside a narrow quartz tube to avoid the sulfurization of the $MoO_3$ before being evaporated. The distances between the precursors and the substrates are indicated by red arrows, while the sizes of zones 2 to 4 of the furnace were 18 cm.

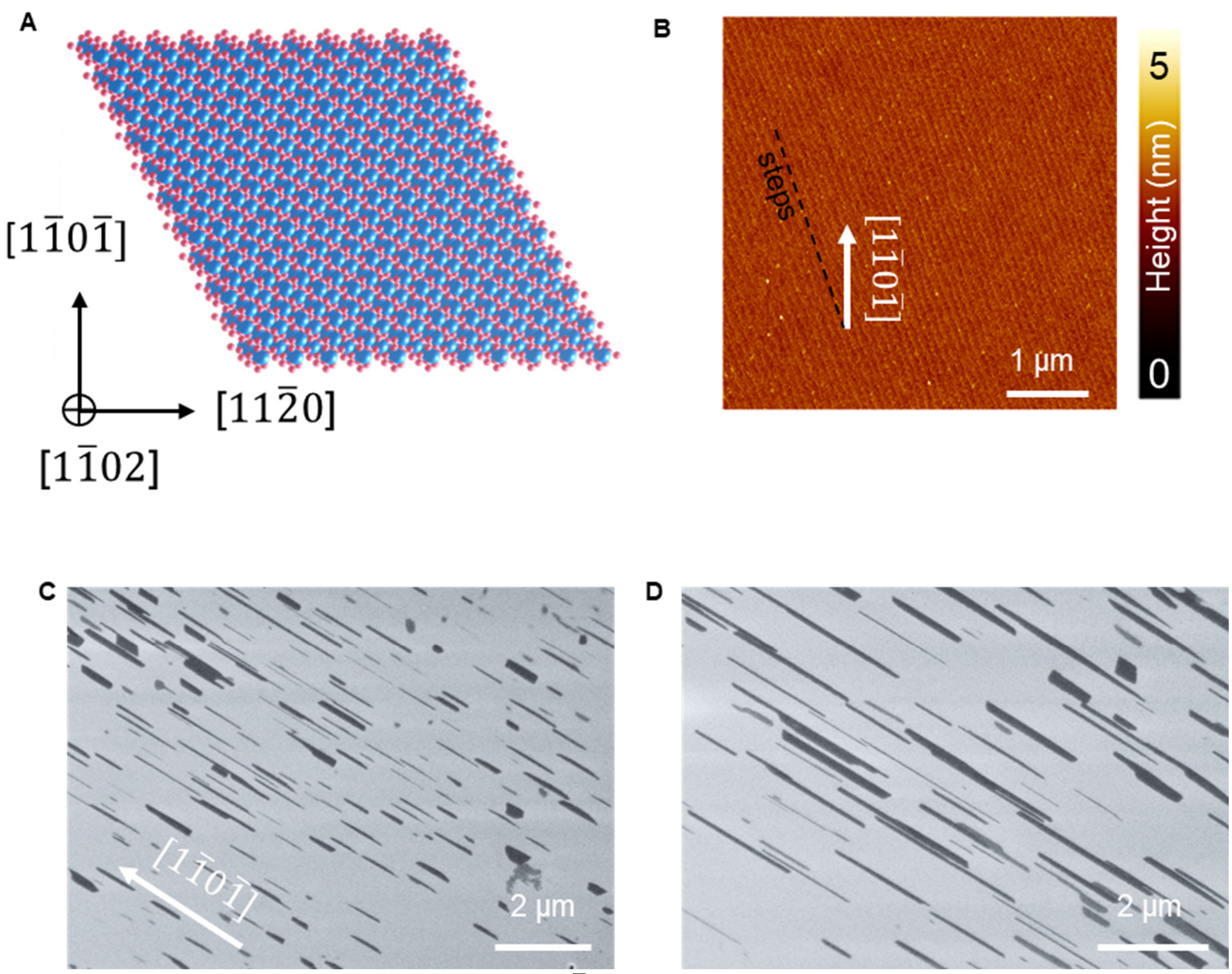

**Fig. S13**. Growth of $MoS_2$ NRs on r-plane sapphire (1$\bar{1}$02). (**A**) Crystal lattice of the r-plane of sapphire. (**B**) AFM image of the surface of r-sapphire, showing the presence of atomic steps. The steps are not parallel to the [1$\bar{1}$0$\bar{1}$] direction of the sapphire. (**C-D**) SEM images showing the presence of $MoS_2$ NRs on the r-plane sapphire, oriented along the [$1\bar{1}0\bar{1}$] direction.

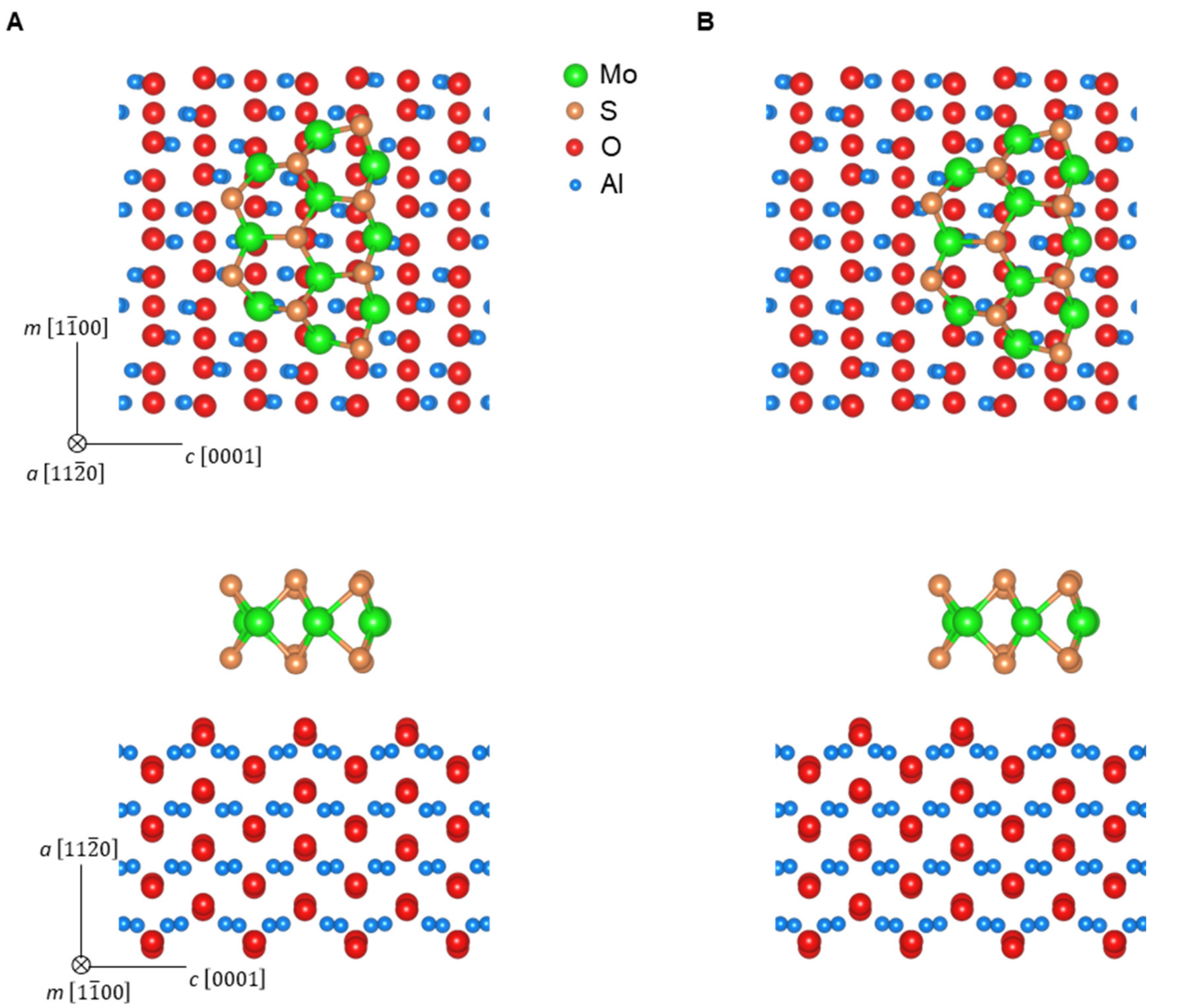

**Fig. S14**. Atomic models for the most and least stable DFT configurations. (**A-B**) Atomic models used for the DFT calculations for the most (**A**) and the least (**B**) stable configurations of the $MoS_2$ on the a-plane sapphire. The models are seen from the [*11$\bar{2}$0*] (top) and [*1$\bar{1}$00*] (bottom) directions of the sapphire lattice, and include the whole sapphire slab used in the calculations.

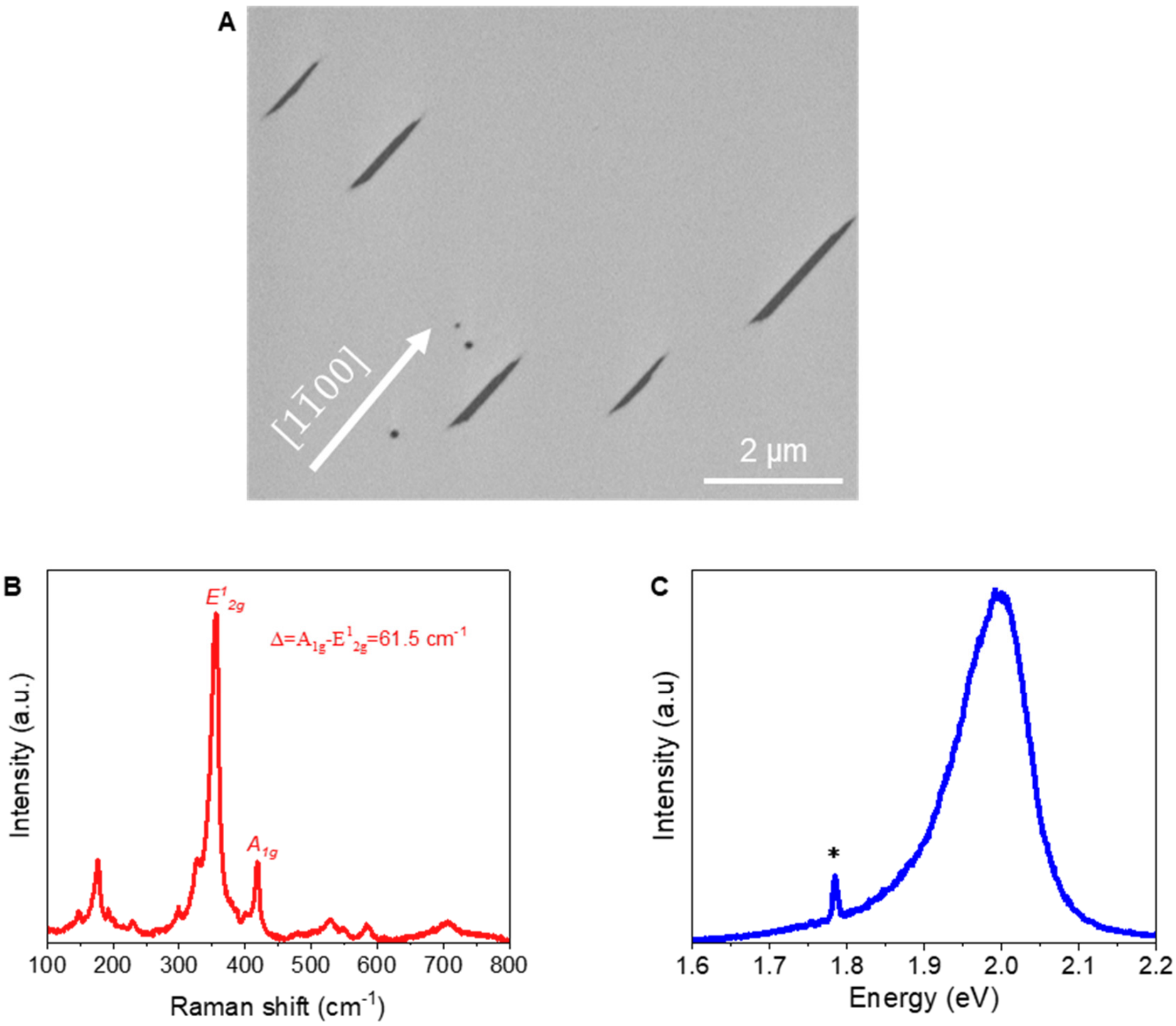

**Fig. S15**. Optical characterization of $WS_2$ nanoribbons. (**A-C**) SEM image (**A**), Raman (**B**) and PL (**C**) spectra of $WS_2$ NRs grown on a-plane sapphire by changing the precursor from $MoO_3$ to $WO_3$. The signal from the sapphire is marked with an asterisk in (C).

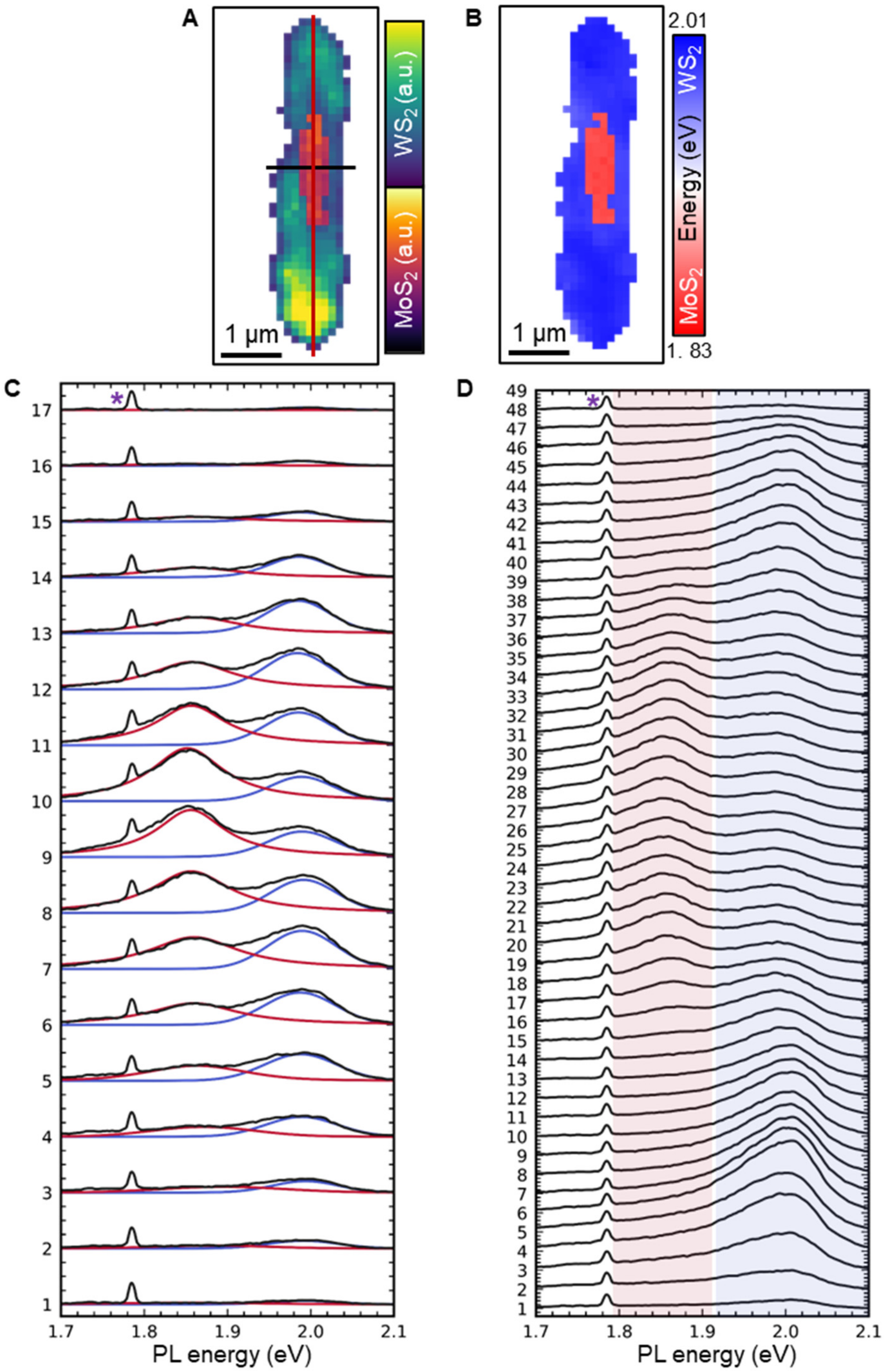

**Fig. S16**. PL characterization of $MoS_2$-$WS_2$ hetero-nanoribbons. (**A-B**) PL intensity (**A**) and energy (**B**) mapping images of a $MoS_2$-$WS_2$ hetero-nanoribbon. (**C-D**) Spectra collected along the horizonal (**C**) and vertical (**D**) lines indicated in (A). The spectra in (C) include the original spectra (black), and the fitted components for $MoS_2$ (red) and $WS_2$ (blue). For clarity, only the original spectra are shown in (D). Spectra in both (C) and (D) were collected at regular intervals of 100 nm along the corresponding lines in (A). The peak highlighted by the asterisk in (C) and (D) originates from the sapphire substrate.

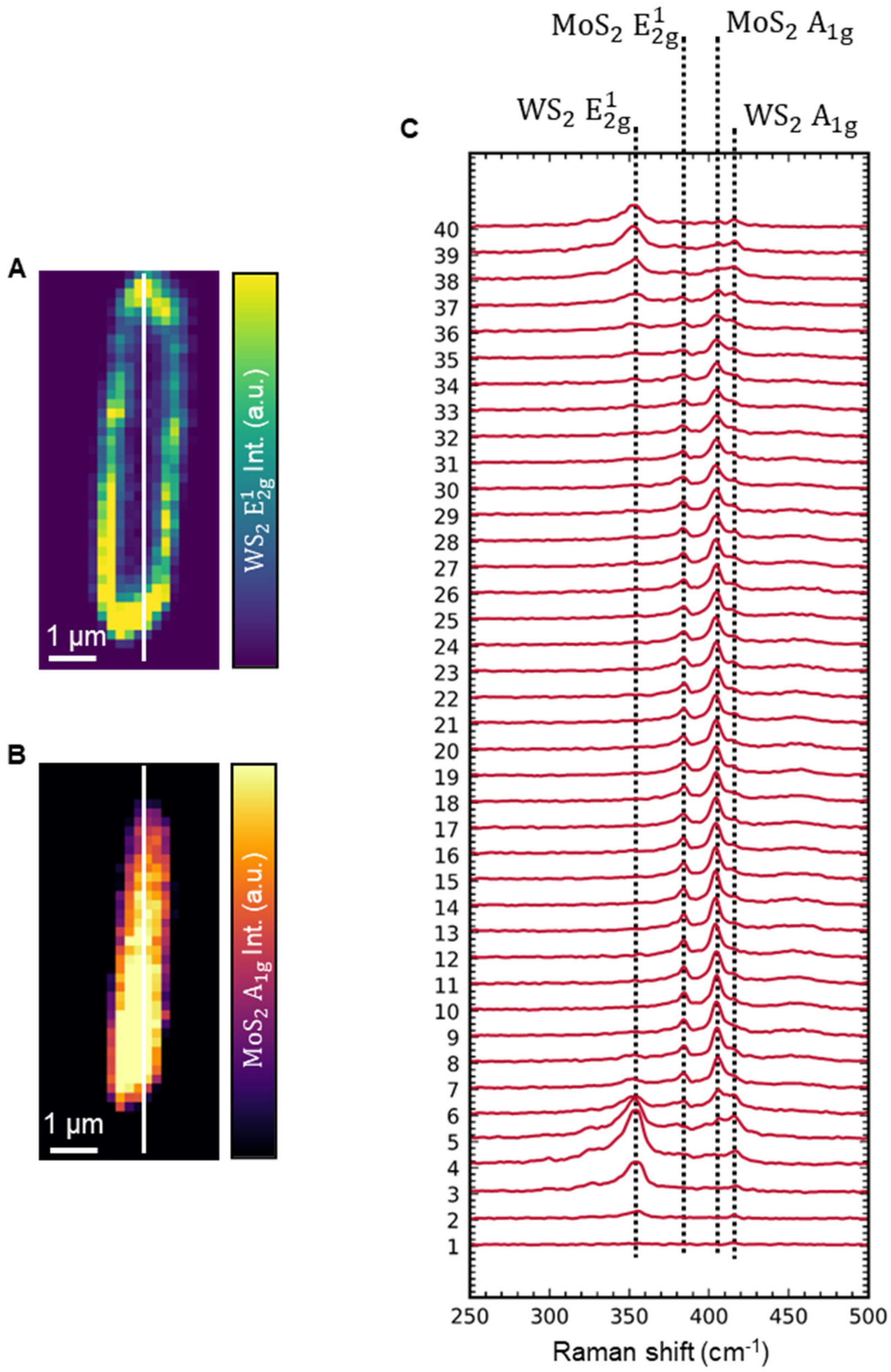

**Fig. S17.** Raman characterization of $MoS_2$-$WS_2$ hetero-nanoribbons. (**A-B**) Mappings of the intensity of the (**A**) $WS_2$ $E^1_{2g}$ and (**B**) $MoS_2$ $A_{1g}$ Raman peaks for a $MoS_2$-$WS_2$ hetero-nanoribbon. **(C)** Raman spectra collected along the white line in (A) and (B), showing the spatial evolution of the $MoS_2$ and $WS_2$ characteristics peaks.

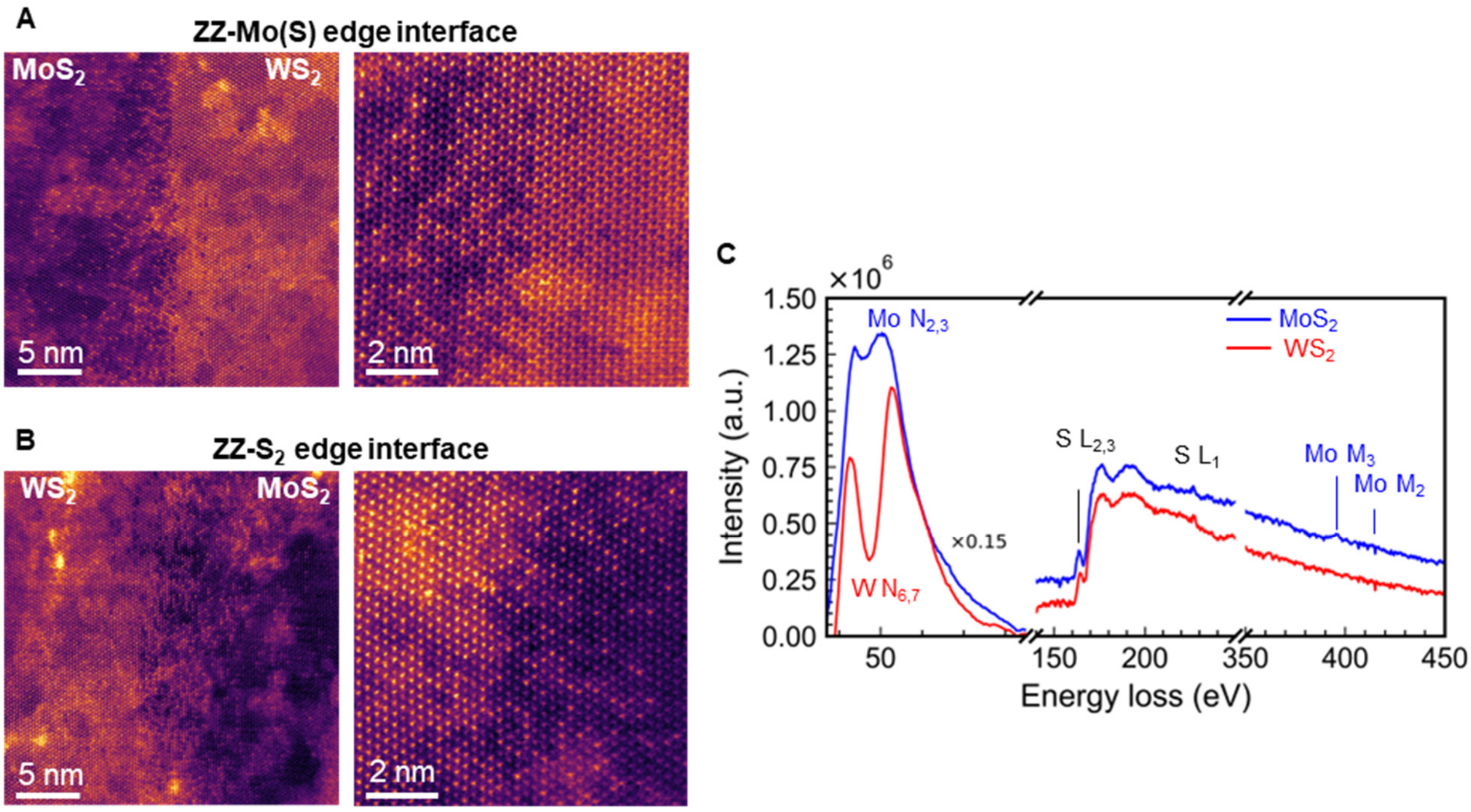

**Fig. S18.** Characterization of the $MoS_2$-$WS_2$ interface of hetero-nanoribbons. (**A-B**) High resolution STEM images of the $MoS_2$-$WS_2$ interface of a hNR at the $MoS_2$ ZZ-Mo(S) (A) and ZZ-$S_2$ (B) termination edges. The interface at the smoother ZZ-Mo(S) edge of the $MoS_2$ has a smaller diffusion region. Brighter and darker atoms correspond to W and Mo, respectively. (**C**) EELS spectra of the $MoS_2$ and $WS_2$ regions of a $MoS_2$-$WS_2$ hNR.

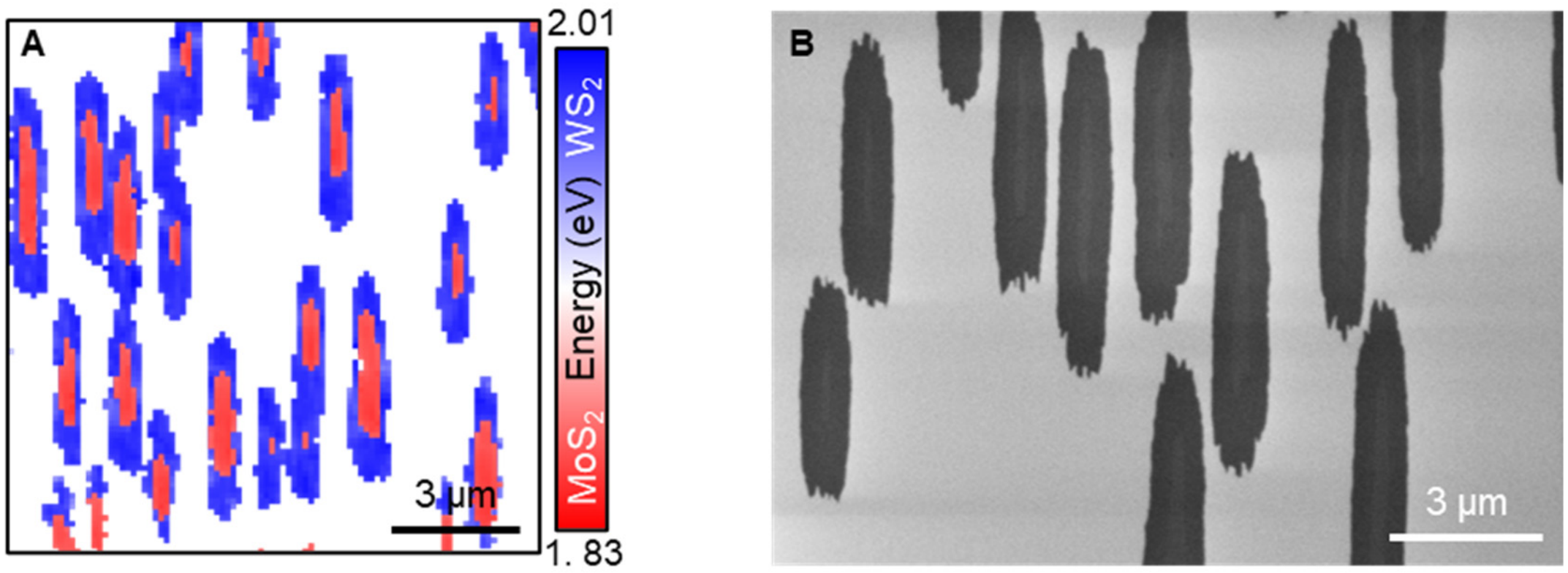

**Fig. S19**. Wide-area characterization of the hetero-nanoribbons. (**A**) Wide-area PL energy mapping of as-grown $MoS_2$-$WS_2$ hNRs, corresponding to the same area as that of Fig. 4H. (**B**) SEM image of as-grown $MoS_2$-$WS_2$ hNRs, showing a contrast between the $MoS_2$ seed at the center, and the $WS_2$ surrounding it.

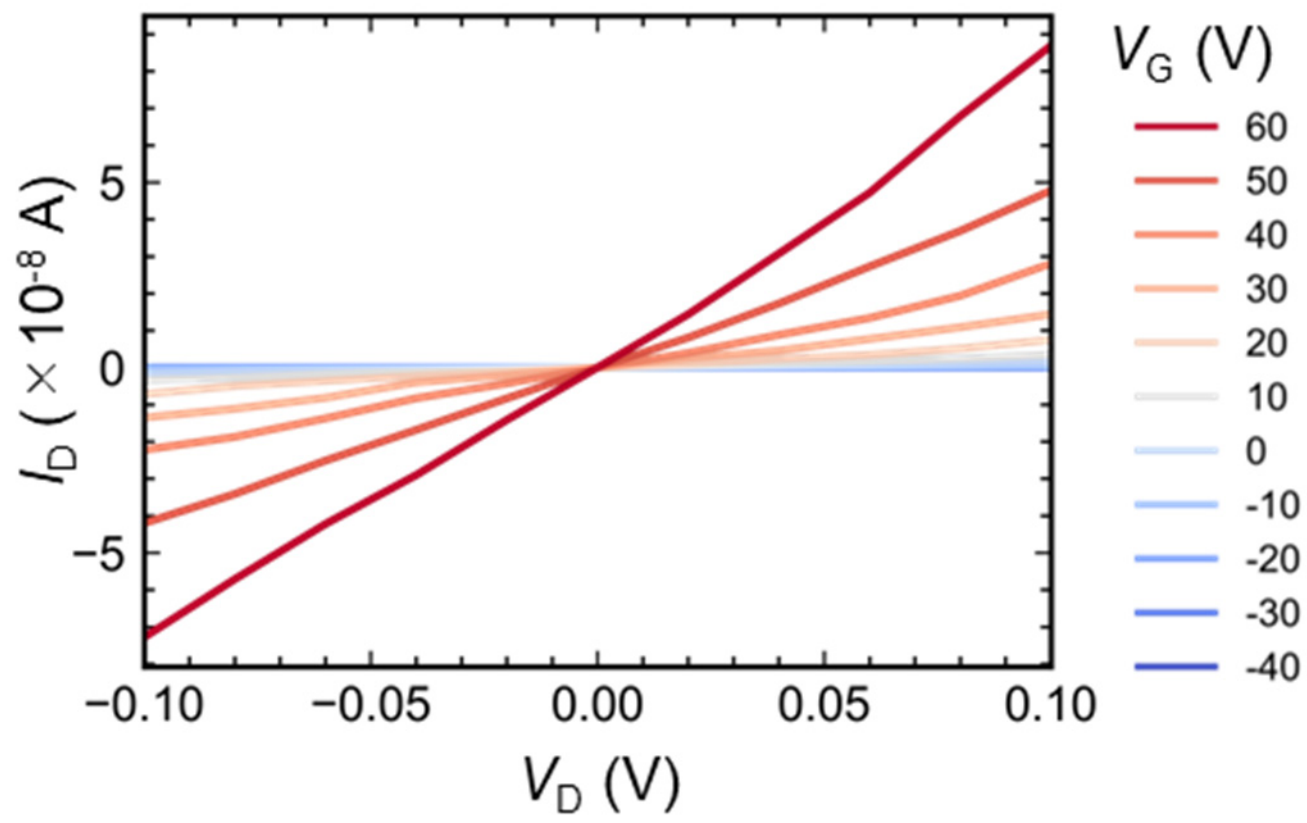

**Fig. S20**. Output characteristics of the $MoS_2$ NR FET of Fig. 6B, measured at room temperature.

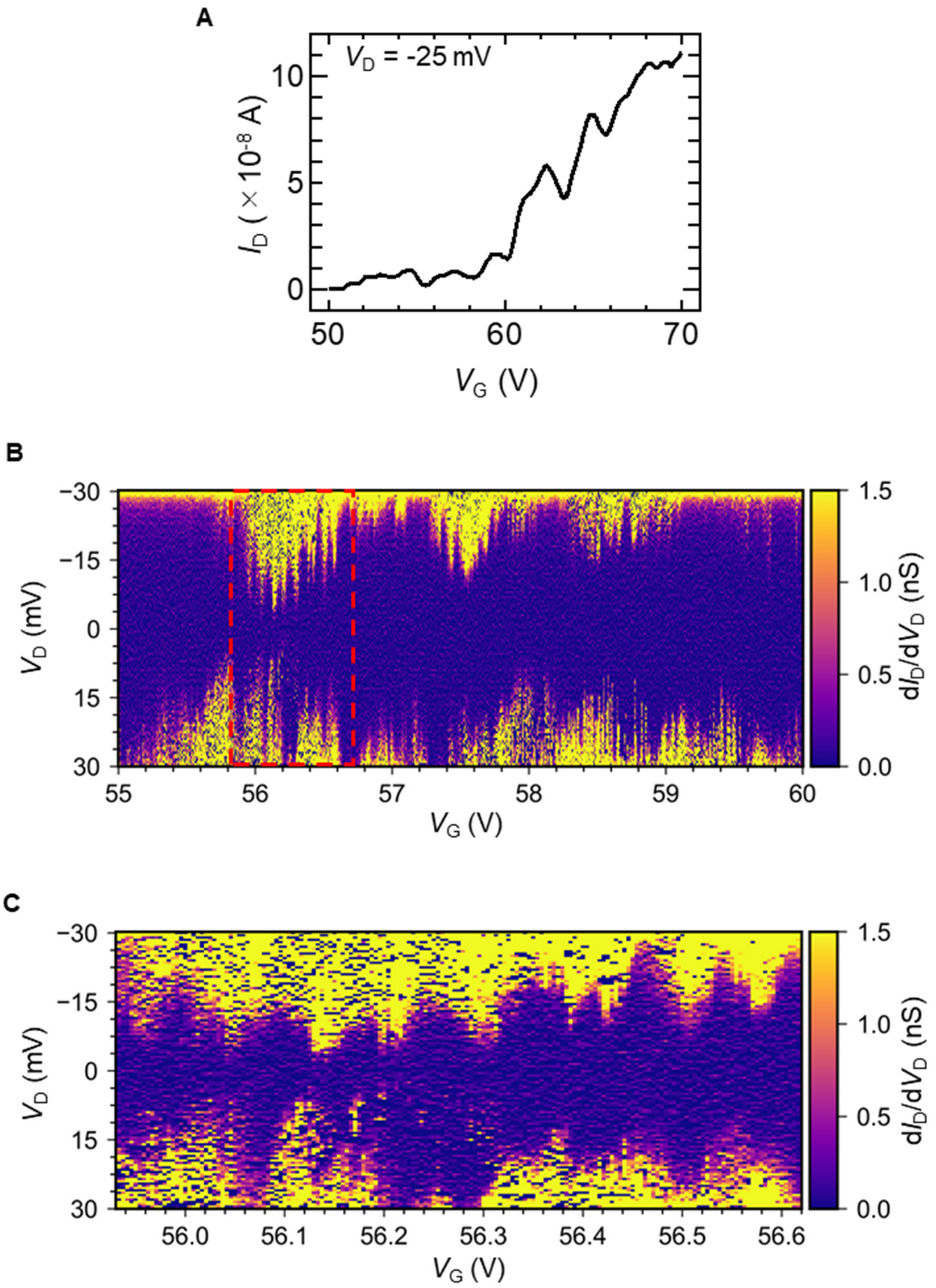

**Fig. S21**. Low-temperature electrical characterization of the $MoS_2$ nanoribbons. (**A-B**) Typical low-temperature (4.2 K) transfer characteristics (**A**) and wide-range Coulomb diamond features (**B**) of a device with a NR channel of width 30 nm. (**C**) Enlargement of the area highlighted in (B), showing the short-period Coulomb diamond features of the device.

**Table S1**.

| | **Tafel slope by SECCM (mV/dec)** | **Tafel slope by bulk (mV/dec)** |
|---|---|---|
| **$MoS_2$ NR** | 139 (current results) | 93 (*59*) |
| **$MoS_2$** | 130 (edge) (*48*) | 110 (*60*) |
| **$MoSe_2$** | Not found in the literature | 141 (*61*) |
| **$WS_2$** | 185 (terrace) (*48*) | 139 (*62*) |
| **$WSe_2$** | Not found in the literature | 102 (*62*) |
| **Pt** | 105 (*63*), 40 (*64*) | 30 (*60*), 30 (low overpotential) (*65*), 120 (high overpotential) (*65*) |
| **Pd** | Not found in the literature | 30 (*65*), 121 (*66*) |

Comparison of Tafel slope values measured by SECCM and by bulk measurement, for $MoS_2$ NRs, flakes of other transition metal dichalcogenides, and for Pt and Pd. The numbers in parenthesis are the corresponding references from which the values were obtained.

**Movie S1**.

The total energy of a trapezoidal-shaped NR-like $MoS_2$ flake on top of a-plane sapphire, computed across a range of relative positions of the flake over the sapphire lattice. The flake is moved along the [$1\bar{1}00$] direction with a low energy barrier (left side) and along the [*0001*] direction with a high energy barrier (right side).